\documentclass[a4paper]{ar-1col}

\usepackage{amsmath,amsthm,amssymb,color,natbib,animate,subcaption}
\newcommand{\bU}{\mathbf{U}}
\newcommand{\bu}{\mathbf{u}}
\newcommand{\bX}{\mathbf{X}}
\newcommand{\bx}{\mathbf{x}}
\newcommand{\bY}{\mathbf{Y}}
\newcommand{\by}{\mathbf{y}}
\newcommand{\bm}{\mathbf{m}}
\newcommand{\N}{\mathcal{N}}
\newcommand{\U}{\mathcal{U}}
\newcommand{\Nat}{\mathbb{N}}

\newcommand{\E}{\mathbb{E}}
\newcommand{\pr}{\mathbb{P}}
\newcommand{\eps}{\boldsymbol{\varepsilon}}
\newcommand{\btheta}{\boldsymbol{\theta}}

\begin{document}

\markboth{Fearnhead and K\"unsch}{Particle Filters}

\title{Particle Filters and Data Assimilation}

\author{Paul Fearnhead$^1$ and Hans R. K\"unsch$^2$
\affil{$^1$Department of Mathematics and Statistics, Lancaster University,
  Lancaster, U.K., LA1 4YF; email: p.fearnhead@lancaster.ac.uk}
\affil{$^2$Seminar f\"ur Statistik, ETH Zurich, Zurich, Switzerland,
  CH-8006; email: kuensch@stat.math.ethz.ch}}

\begin{abstract}
State-space models
can be used to incorporate subject knowledge on the underlying dynamics of
a time series by the introduction of a latent Markov state-process. A user
can specify the dynamics 
of this process together with how the state relates to partial and noisy observations that have been made. Inference and prediction
then involves solving a challenging inverse problem: calculating the conditional distribution of quantities of interest given the observations. This article
reviews Monte Carlo algorithms for solving this inverse problem,
covering methods based on the particle filter and the ensemble Kalman
filter. We 
discuss the challenges
posed by models with high-dimensional states, joint estimation of
parameters and the state, and inference for the history of the state
process. We also point out some potential new developments which will be
important for tackling cutting-edge filtering applications.

\end{abstract}

\maketitle


\section{INTRODUCTION}
This article gives an overview of Monte Carlo methods for
estimating parameters and latent variables and for making predictions 
in state space models. 
 In some fields state space models are known by the name
of hidden Markov models; but we will use the term state space model throughout.

\subsection{What Is a State Space Model?}
In order to predict a time series of observations, it is essential
to take subject knowledge of the dynamics of the series into account.
However, in many applications subject knowledge involves adding
to the observed variables other variables that are hard or impossible
to measure. A state space model specifies the joint
distribution of all the variables that are required for a dynamical
model based on subject knowledge, and the variables that have been
observed. The former are called the state variables and are denoted
by $(\bX_t; t \geq 0)$. The evolution of the state variables is assumed to 
be given either by a Markov process or deterministically by a system of
ordinary or partial differential equations. The state variables are
latent; we only have access to observations
$(\bY_i; i \in \Nat)$ that are partial and noisy functions 
of the state $\bX_{t_i}$ at observation times $t_i$.

In some applications, the state variables are not obtained by a detailed
subject-based modeling, but rather represent dynamic random effects
or unknown  time-varying parameters that have a simple dynamics, often a
linear Gaussian autoregression. Combined with a generalized
linear model for the observations given the states, this leads to
what \cite{Cox:1981} calls parameter-driven models. 

\subsubsection{Example 1: Tracking} The particle filter methods we are reviewing
in this article were first motivated by tracking applications \cite[e.g.][]{Gordon/Salmond/Smith:1993,stone2013bayesian}.
For these applications the state will be the position and velocity of the target or targets
being tracked. Observations are made of their location, but can be partial (only measurement of the bearing), noisy,
and can include clutter (spurious measurements that do not relate to any target). For these applications the key
inference questions relate to estimating the current positions of targets and predicting their future movement. This requires
on-line algorithms, such as particle filters, that can quickly update
beliefs of the state as each new measurement is observed.


\subsubsection{Example 2: Numerical Weather Prediction} Advances in
numerical weather prediction during the past 50 or 100 years have been
termed  a ``quiet revolution'' by \cite{Bauer:15} in a ``computational
problem comparable to the simulation of the human brain and of the evolution of
 the early Universe.'' These advances have been made possible not 
only by increased computing power, better measurements and improved 
physical understanding, but also by ensemble forecasts, which quantify
uncertainty, and data
assimilation methods which sequentially integrate measurements 
into the forecasting process.

\subsubsection{Example 3: Ecology} A model for the evolution of a
population usually needs information about the abundance in different
age classes. The dynamics of the model relate abundances
at the next time-point to current abundances whilst accounting for rates
of fertility, mortality, catchment and migration.
The models thus have states which record population sizes within each age
range, and those rates that are considered time-varying.  
Observations,
for example from capture-recapture experiments,
will relate indirectly to these population sizes. Interest is often about future predictions about
the population, which requires estimates of both the current state and the parameters.
For more details see \cite{Flemming:17} or \cite{Nielsen:2014}.

\subsection{What Are Filtering and Data Assimilation?}

In order to apply state space models, we need to be able to 
estimate unobserved states, future observations and unknown
parameters of the model from available data. For this, the
key task is to compute the conditional distribution of
the state $\bX_{t_i}$ at time $t_i$ based on observations up to time
$t_i$, the so-called filtering distribution. Once we know
this filtering distribution, we can obtain predictive
distributions of future states by letting the state process
evolve with the filtering distribution as initial distribution at 
time $t_i$. From the predictive distribution of future states,
the predictive distribution of observations follows immediately.
All the relevant information about future states and observations
is thus contained in the filtering distribution.

Computing the filtering distribution is, however, a difficult task.
Some simplification occurs by exploiting a recursive scheme.
Using the filtering distribution at time $t_{i-1}$, we first
compute the predictive distribution of $\bX_{t_i}$ based on
observations up to time $t_{i-1}$, using the dynamics of the
state. The filtering distribution at time $t_i$ then follows 
from Bayes' formula applied with the predictive distribution
as the prior and using  the likelihood of $\bx_{t_i}$ given $\by_i$.
Hence recursive filtering proceeds by an alternation of prediction
or propagation steps based on the dynamics of the state, 
and update steps based on the most recent observation. 

Filtering is engineering terminology. In geophysics the term
data assimilation is used instead. The predictive distribution of
$\bX_{t_i}$ given observations up to time $t_{i-1}$ 
 is usually called the background
distribution, and the filter distribution is usually called
the analysis distribution. The exchange of ideas and methods
for filtering between statistics
on the one hand and geophysics and applied mathematics on the other
has only recently become more common, and one aim of this review
is to bring the two communities closer together. 

In geophysics, the state evolution is often deterministic, but
chaotic, i.e., sensitive to initial conditions. In fact, the
phenomenon of chaos was discovered in a toy atmospheric physics
model by \cite{Lorenz:63}. Because of this sensitivity
to initial conditions, new observations have to be assimilated
frequently for good predictions. 

Except in special cases, the propagation and the update steps cannot
be computed analytically. As the steps involve integration over
the, often high-dimensional, state space, Monte Carlo
approximations are currently preferred. This review
is limited to these approximations.  

 
\subsection{Outline  of the Review}

After giving some background on state space models and a brief treatment of
the basic recursions for the true filtering and predictive distributions in
Section \ref{S:FilterRecursions}, we describe in Section
\ref{S:MonteCarloAlgorithms} 
Monte Carlo methods to approximate these recursions, namely the 
particle filter, the ensemble Kalman filter and their extensions.
Section \ref{S:Convergence} briefly summarizes theoretical properties of
the particle filter, and Section \ref{S:Collapse} discusses
the challenges that arise when applying these filter methods to models with
high-dimensional states.


We then focus on methods for smoothing and parameter estimation. Smoothing involves calculating the conditional
distribution of historic values of the state given all observations to date. We show how
particle filter ideas can be extended and applied to approximate these smoothing distributions in Section \ref{S:Smoothing}. Then in Section
\ref{S:ParameterEstimation} we look at particle filter methods for estimating parameters, with particular emphasis on recent
particle MCMC methods. The review ends with a summary and outlook.

We do not make an attempt to give a complete overview of all aspects of
filtering or to provide a comprehensive list of references. Recent other
reviews are \cite{Doucet:11}, \cite{Kunsch:13} and \cite{Kantas:2015},
while part III of \cite{Douc:14book} contains a detailed introduction 
with many examples and proofs. 
 \cite{Majda:12} and \cite{Reich:15} present the field from
applied mathematics and geophysics perspectives. One area we view as important, but do not cover,
is the increasing need to design filtering algorithms that can take advantage of modern computer architecture.
This is an area we flag later as an important future issue, but see \cite{lee2015forest} and \cite{verge2015parallel} for some recent work.

Software, in {\texttt{R}}, for implementing some of the examples we consider in this paper are available as online supplementary material. This is provided
primarily to give the reader the opportunity to run the algorithms under
different settings so as to build up a stronger intuition as to when and
why any of these methods work well. Software for implementing some of the methods
we describe in this review for generic applications is also available.
We are aware of the following: {\texttt{SMCTC}} \cite[]{johansen2009smctc},
{\texttt{LiBi}} \cite[]{Lawrence:2015}, the package {\texttt{nimble}}
\cite[]{Michaud:17} in {\texttt{R}, the Robotics System Toolbox of
  {\texttt{MATLAB}}, and {\texttt{PDAF}} \cite[]{nerger2013software} and
  {\texttt{DART} (see \verb+http://www.image.ucar.edu/DAReS/DART/+ ) for  
geophysical applications.

\section{STATE SPACE MODELS AND FILTER RECURSIONS} \label{S:FilterRecursions}


\subsection{State Space Models}

In order to simplify the notation, we assume that the observation
times are equally spaced with $t_i=i$ and that the model is
time-homogeneous. All results and
methods can be easily extended to unequally spaced 
observations and time-inhomogeneous models.  We also repeatedly use the
notation that the subscript $s\!\!:\!\!t$ refers to the set of values at all
times from time $s$ to time $t$, so, for example, $X_{0:n}=(X_0, \ldots, X_n)$. 

The state process $(\bX_t)$ is assumed to be Markovian and the $i$-th
observation, $\bY_i$, depends only on 
the state at time $i$, $\bX_i$, and is conditionally independent of all other
observations. This means that
\begin{eqnarray}
\bX_t \mid  (\bx_{0:t-1},\by_{1:t-1}) &\sim& P(d\bx_t \mid \bx_{t-1}), 
\quad \bX_0 \sim \pi_0(d\bx_0) \label{eq:state-trans}\\
\bY_t \mid (\bx_{0:t},\by_{1:t-1}) &\sim& g(\by_t\mid \bx_t) d\nu(\by_t). 
\label{eq:obs-dens}
\end{eqnarray}
If the state evolution is
given by a time-homogeneous (autonomous) differential equation, 
$P$ becomes a point mass at 
the solution at time $t$ with initial conditon $\bx_{t-1}$ at time
$t-1$. Similarly, other common models, such as state-space formulations of AR($p$) models when $p\geq 2$, can mean that components of $\bX_t$
are deterministic functions of $\bX_{t-1}$.
We therefore do not want to assume that the state transitions
have densities, but the conditional 
distribution of the observations should have densities so that we can
use Bayes' formula (though extensions of filters to exact observation of part of the state is possible, see Section \ref{S:AuxiliaryParticleFilter}).
The measure $\nu$  is usually either Lebesgue or
counting measure. In practice either or both of the distributions that determine the state evolution
or the measurement process can depend on parameters. Many particle filter methods assume such
parameters are known. We will suppress the dependence of $P(d\bx_t\mid  \bx_{t-1})$ and $g(\by_t\mid \bx_t)$
on such parameters in our notation except when we consider estimating the parameters in
Section \ref{S:ParameterEstimation}.


State space models are directed graphical models \citep{Lauritzen:96} 
with the following graph
\begin{center}\begin{tabular}{ccccccccc}
$\ldots$&$\rightarrow$&$\bX_{t-1}$
&$\rightarrow$& $\bX_{t}$ &$\rightarrow$& $\bX_{t+1}$& $\rightarrow$& \ldots \\
&& $\downarrow$&  &$\downarrow$& &$\downarrow$&& \\
$\ldots$&&$\bY_{t-1}$&&$\bY_{t}$ && $\bY_{t+1}$ && \ldots \\
\end{tabular}\end{center}
Various conditional independence properties
follow from this graph. For instance, $\bX_t$ is conditionally independent of 
$(\bY_{t+1}, \bX_{t+2}, \bY_{t+2}, \ldots)$ given $\bX_{t+1}$. Such properties will
be used in Section  \ref{S:Smoothing}.


\subsubsection{Two examples}  


\paragraph{Example: Stochastic Volatility} To help demonstrate the different 
algorithms clearly in the figures and animations, we will use an example
with a 1-dimensional state:
\[
 X_t\mid x_{t-1} \sim \N\left(\phi x_{t-1},\sigma^2\right), ~~~ Y_t\mid x_t \sim \N\left(0,\beta^2\exp\{x_t\}\right).
\]
This is a simple stochastic volatility model \cite[see e.g.][]{kim1998stochastic}, with the state, $X_t$, being proportional to the log-volatility
of the observation series. The model has three parameters, $\phi$, $\sigma$ and $\beta$ which respectively 
govern the dependence and noise in the state process, and the base-line variance of the observation process.

\paragraph{Example: Lorenz 96} This is a toy model of a one-dimensional
atmosphere, popular as a test bed for data assimilation in atmospheric
physics \citep{Lorenz:96}. In the supplemental material we will use it
to illustrate the ensemble Kalman filter. 
The state is 40-dimensional with dynamics given by the
differential equation 
\begin{equation*}
  \frac{dX_{t,k}}{dt} = (X_{t,k+1} - X_{t,k-2}) X_{t,k-1} - X_{t,k} + 8, \quad
  k=1, \dots, 40, \ X_{t,k} \equiv X_{t,k+40}.
\end{equation*}
At times $t=i\Delta$, every $m$-th component of $X_t$ is observed with
independent additive Gaussian noise. This can be written as
$\bY_t\mid  \bx_{t} \sim \N(H \bx_{t}, \sigma^2 I)$
where $H$ is the appropriate matrix to  select the observed components.

\subsection{Prediction, Filter and Smoothing Distributions}
We collect here the basic formulae of
conditional distributions and likelihoods that are needed for prediction,
filtering, smoothing and parameter estimation. 

For $0 \leq s \leq u$ and $t \geq 1$, the conditional distribution
of $\bX_{s:u}$ given $\bY_{1:t}=\by_{1:t}$ is denoted by  
$\pi_{s:u| t}$, and we use $\pi_{s | t}$ instead of $\pi_{s:s|t}$. Hence 
$\pi_{t| t-1}$ is the predictive distribution at time $t$ based on
observations up to time $t-1$, for short the prediction
distribution at time $t$. Finally we denote
the filtering distribution at time $t$ by $\pi_t$ instead of $\pi_{t| t}$.

With a slight abuse of notation, all other (conditional) densities are 
denoted by $p$: the arguments of $p$ indicate which random variables 
are involved.

The assumptions (\ref{eq:state-trans}) and (\ref{eq:obs-dens}) 
imply the following joint distributions 
\begin{equation}
\label{eq:joint}
(\bX_{0:s},\bY_{1:t}) \sim \pi_0(d\bx_0) \prod_{i=1}^s P(d\bx_i\mid \bx_{i-1}) 
\prod_{j=1}^t g(\by_j\mid \bx_j) \nu(d\by_j), ~~ s\geq t.
\end{equation}
Integrating out the path of the state process, we obtain that
$\bY_{1:t} \sim p(\by_{1:t}) \prod_j \nu(d\by_j)$, where
$$p(\by_{1:t}) = \int\pi_0(d\bx_0) \prod_{i=1}^s P(d\bx_i\mid \bx_{i-1}) 
\prod_{j=1}^t g(\by_j\mid \bx_j).$$
 If the model contains unknown parameters $\btheta$,
$p(\by_{1:t})$ becomes  the likelihood of $\btheta$.
By Bayes' formula, $\pi_{0:s\mid t}$ is the
right-hand-side of (\ref{eq:joint}) divided by $p(\by_{1:t})$.  From this
it is easy to see that the following recursion hold:
\begin{eqnarray}
\pi_{0:t| t-1}(d\bx_{0:t}\mid \by_{1:t-1}) &=& 
\pi_{0:t-1\mid t-1}(d\bx_{0:t-1}\mid \by_{1:t-1}) P(d\bx_t \mid \bx_{t-1}) \label{eq:prop1},\\
\pi_{0:t| t}(d\bx_{0:t}\mid \by_{1:t}) &=& 
\pi_{0:t| t-1}(d\bx_{0:t}\mid \by_{1:t-1}) \frac{g(\by_t \mid \bx_t)}{p(\by_t\mid \by_{1:t-1})},
\label{eq:update1}
\end{eqnarray}
where 
\begin{equation}
  \label{eq:likeli}
  p(\by_t\mid \by_{1:t-1}) = \frac{p(\by_{1:t})}{p(\by_{1:t-1})} = \int
  \pi_{t| t-1}(d\bx_{t}\mid \by_{1:t-1}) g(\by_t\mid \bx_t). 
\end{equation}
Integrating out the states $\bx_{0:t-1}$ in (\ref{eq:prop1}) and 
(\ref{eq:update1})  leads to the recursion discussed
in the Introduction:
\begin{eqnarray}
\pi_{t| t-1}(d\bx_{t}\mid \by_{1:t-1}) &=& 
\int \pi_{t-1}(d\bx_{t-1}\mid \by_{1:t-1}) P(d\bx_t\mid \bx_{t-1})\label{eq:prop2},\\
\pi_{t}(d\bx_{t}\mid \by_{1:t}) &=& 
\pi_{t| t-1}(d\bx_{t}\mid \by_{1:t-1}) \frac{g(\by_t\mid \bx_t)}{p_t(\by_t\mid \by_{1:t-1})}.
\label{eq:update2}
\end{eqnarray}
Both recursions consist of
a propagation step,  (\ref{eq:prop1}) or (\ref{eq:prop2}), 
and an update or correction step, (\ref{eq:update1}) or (\ref{eq:update2}). Making predictions more than
one time-step ahead is simple, as we can apply the propagation update (\ref{eq:prop2}) 
without the correction step
\[
 \pi_{t+s\mid t}(d\bx_{t+s}\mid \by_{1:t}) = \int
 \pi_{t+s-1\mid t}(d\bx_{t+s-1}\mid \by_{1:t}) P(d\bx_{t+s}\mid \bx_{t+s-1}), \mbox{ for $s=1,\ldots,$}.
\]
Because the observations, $\by_t$, are fixed,
we will often drop them from the notation, for example writing $\pi_{t}(d\bx_t)$ rather
than $\pi_t(d\bx_t\mid \by_{1:t})$.



\section{MONTE CARLO FILTER ALGORITHMS} \label{S:MonteCarloAlgorithms}

Monte Carlo filter algorithms approximate the filter
distributions, $\pi_{t}$, by weighted samples 
$(\bx^i_t, w^i_t)$ of size $N$:
\begin{equation}
 \pi_{t}(d\bx_t) \approx \hat{\pi}^N_{t}(d\bx_t) =
\sum_{i=1}^N w^i_t \delta_{\bx^i_t}(d\bx_t). \label{eq:filt-appr} 
\end{equation}
Here $\delta_{\bx}$ denotes the point mass at $\bx$. When we simultaneously consider
approximations of 
$\pi_{t}$ by weighted and unweighted samples, we denote the latter
by $(\tilde{\bx}^i_{t})$.  The sample members are called particles because
in the algorithm they will move in space and have offspring or die. 
In geophysics, samples are usually called ensembles.

\subsection{The Bootstrap Filter} \label{S:BootstrapFilter}

If we insert the approximation (\ref{eq:filt-appr}) at time $t-1$ into the
propagation (\ref{eq:prop2}), we obtain
$$ \pi_{t| t-1}(d\bx_t)  \approx \sum_{i=1}^N w^i_{t-1}
P(d\bx_t \mid  \bx^i_{t-1}).$$
Therefore, if $\bx^i_t \sim P(d\bx_t\mid \bx^i_{t-1})$ independently for 
$i=1, \ldots, n$, the weighted sample $(\bx^i_t, w^i_{t-1})$ approximates
the prediction distribution $\pi_{t| t-1}$:
$$\pi_{t| t-1}(d\bx_t) \approx \hat{\pi}^N_{t| t-1}(d\bx_t) =
\sum_{i=1}^N w^i_{t-1} \delta_{\bx^i_t}(d\bx_t).$$
Applying the Bayes' update 
(\ref{eq:update2}) to this approximation gives
\begin{equation}
 \pi_{t}(d\bx_t)  \approx \hat{\pi}^N_{t}(d\bx_t) =
\sum_{i=1}^N w^i_t \delta_{\bx^i_t}(d\bx_t),
 \quad w^i_t \propto W^i_t= w^i_{t-1} g(\by_t\mid \bx^i_t). \label{eq:approx-filter}
\end{equation}
This closes the recursion of the sequential importance sampling algorithm. 
At time $t=0$ we initialize it by drawing 
$\bx^i_{0}$ from $\pi_0(d\bx_0)$ and set $w^i_0=1/N$. As a by-product,
the normalizing constant for the weights $w^i_t$ provides
an approximation of $p(\by_t\mid \by_{1:t-1})$ because
$$\E\left(\sum_{i=1}^N W^i_t \mid (\bx^j_{t-1})_{j=1}^N  \right) = 
\sum_{i=1}^N w^i_{t-1} 
\int g(\by_t\mid \bx_t)P(d\bx_t\mid \bx^i_{t-1}) \approx 
\int g(\by_t\mid \bx_t)\pi_{t | t-1}(d\bx_t).$$


The sequential importance sampling algorithm has the drawback
that after a few iterations the weights are essentially concentrated on a 
few particles and most or all particles are in regions where the
true filter distribution has little mass. To avoid this, the basic
bootstrap filter makes weights equal by resampling before propagating.
It consists of the following steps:
\begin{itemize}
\item[(1)] Resample: Set $\tilde{\bx}^i_{t-1} = \bx^{A(i)}_{t-1}$
  where $\pr(A(i)=j) = w^j_{t-1}$ for $i=1, \ldots, N$.
\item[(2)] Propagate: Draw $\bx^i_{t}$ from $P(d\bx_t\mid \tilde{\bx}^i_{t-1})$,
  independently for $i=1, \ldots, N$.
\item[(3)] Reweight: Set $w^i_{t} \propto W^i_t = g(\by\mid \bx^i_{t})/N$.
\item[(4)] Likelihood Estimation: Calculate $\hat{p}(\by_t\mid \by_{1:t-1})=\sum_{i=1}^N W^i_t$, and set $\hat{p}(\by_{1:t})=\hat{p}(\by_{1:t-1})\hat{p}(\by_t\mid \by_{1:t-1})$.
\end{itemize} 
See Figure \ref{Fig:Bootstrap} for an example of the output of this recursion.

The computational complexity of one iteration of the bootstrap filter is $O(N)$.
The observation likelihood, $g$, is required in closed form, whereas
we need only  
to be able to simulate from the propagation distribution, $P$. Particles interact through the normalisation in the reweighting step.

As a by-product of running the bootstrap filter we get an approximation of the prediction distribution
by the unweighted sample $(\bx^i_t)$ in step (2). Step (4) is optional, but gives
an estimate of the parameter likelihood. The estimate $\hat{p}(\by_{1:t})$ is
unbiased, see Theorem 7.4.2 in \cite{del2004feynman} or \cite{Pitt:2012}, 
a property that will be used in Section \ref{S:PMCMC}. However,
$\hat{p}(\by_t\mid \by_{1:t-1})$ is biased in general. 

\begin{figure}
 \begin{center}
    \includegraphics[width=0.9\textwidth]{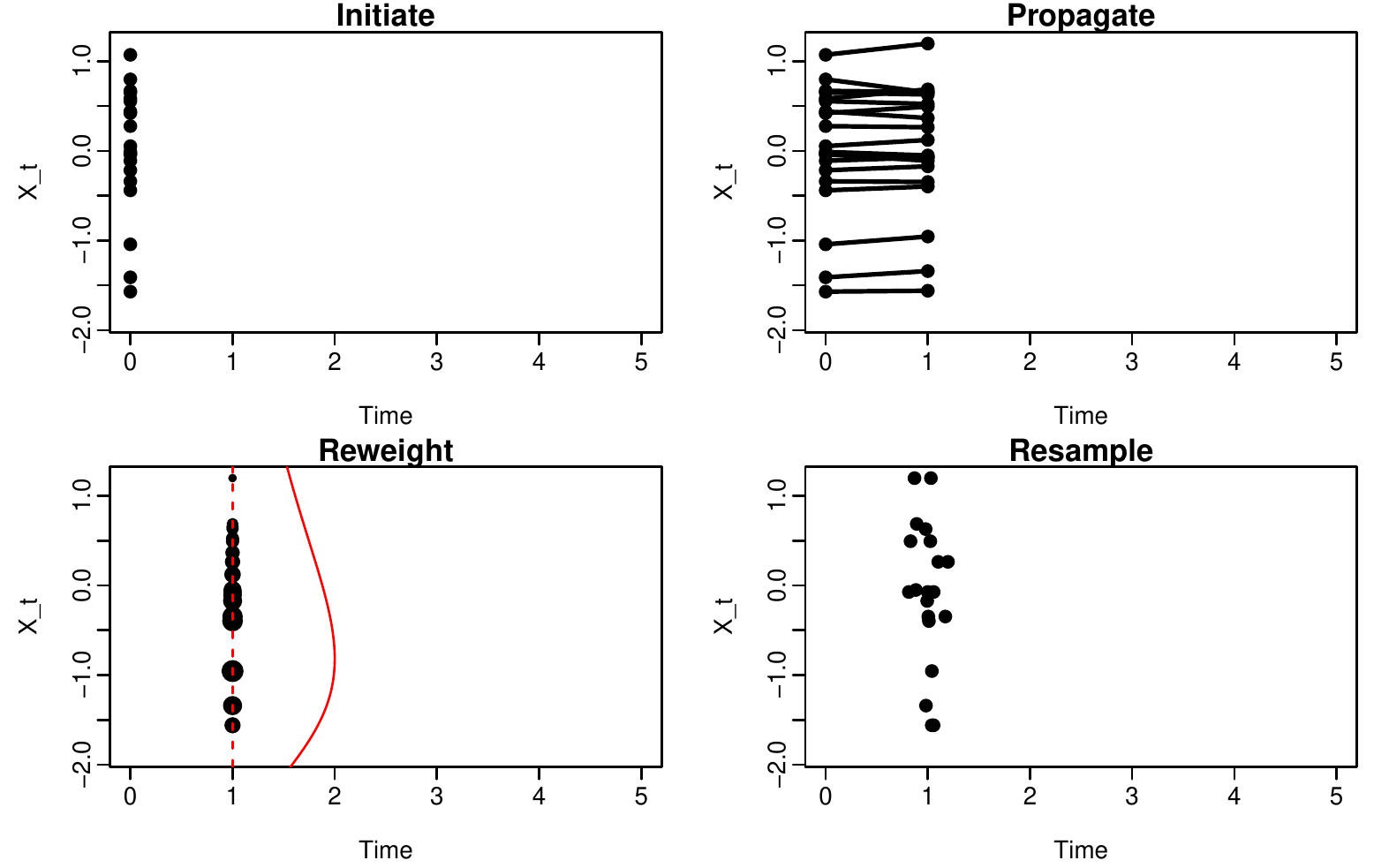}  
 \end{center}   
\caption{\label{Fig:Bootstrap} 
Plots of the bootstrap filter for the stochastic volatility model. We show the output after each of the three stages: propagate, reweight and resample. For the reweight plot we show the likelihood
function, $g(x_t)$, and re-scale the particles according to the weight they are given. For the resample plot we have jittered the time-component of the particles purely so that one can see when multiple
copies of a particle are resampled. An animated version of the figure is
available in the supplemental material.
}
\end{figure}

We call $\bx^{A(i)}_{t-1}$  the ``ancestor'' of particle $\bx^i_t$ and
denote the
number of ``offspring'' of particle $\bx^i_t$  at time $t$ by $N^i_t$.
Resampling replaces the weights $w^i_{t}$
by random  weights $N^i_t/N$ and thus always increases the Monte 
Carlo error of the approximation to the current filtering distribution. 
However resampling is beneficial when we propagate particles to future time-steps. 
It stochastically removes particles with low weight, and produces multiple copies of
particles with high weight.
Multiple copies of a current particle are then able to independently explore the future of the state.

The benefit of resampling depends crucially on the level of stochasticity in the 
propagation distribution,  $P$. If this is high relative to the filter variance, then even if
the current particles lack diversity, this diversity is quickly regained as we propagate forward in time.
If the state dynamics are deterministic we will not regain diversity as we propagate the
particles forward. To overcome this, it is possible to add random noise to the current particles prior
to propagation, which can be justified as sampling from a kernel density approximation to the filter density
\cite[]{Liu/West:2001}. The issue of deterministic dynamics arises not
only in many geophysical applications, but also 
when we have unknown, fixed parameters in the model. We discuss this issue in more detail in Section \ref{S:ParameterEstimation}.

The ancestors $A(i)$ do not have to be drawn independently for different
$i$; it is only required that  $\E(N^i_t) = N w^i_{t}$.
Balanced sampling \cite[also called stratified or systematic sampling][]{Kitagawa:1996,Carpenter/Clifford/Fearnhead:1999} makes the resampling 
error as small as possible. 
The simplest method partitions the interval $[0,N]$ into subintervals
of length $N w^i_t$ and counts how many points of the sequence
$U,U+1, \ldots, U+N-1$ where $U \sim \U(0,1)$ fall in each subinterval.
See \cite{Crisan:2001}  for a different balanced sampling method.

If the weights, that is the values of the observation likelihood, are very
unbalanced, 
resampling risks losing too much diversity that cannot be restored
correctly in the propagation step, even with stochastic dynamics.  
This problem, and ways to alleviate or overcome it, will be 
considered in the rest of this section and in Section \ref{S:Collapse}.

\subsection{Auxiliary Particle Filters} \label{S:AuxiliaryParticleFilter}
The bootstrap filter uses importance sampling for a target 
proportional to $\sum_i P(d\bx_t\mid \tilde{\bx}^i_{t-1}) \cdot g(\by_t\mid \bx_t)$
with the proposal $N^{-1}\sum_i  P(d\bx_t\mid \tilde{\bx}^i_{t-1})$. If the 
observation likelihood
is informative, the target and the proposal are not close enough and
the weights become unbalanced. In this case, we can try to find a better
proposal, such as $ \sum \tilde{w}_{t-1}^{(i)}\tilde{P}(d\bx_t\mid \tilde{\bx}^i_{t-1})$ where the weights, $\tilde{w}_{t-1}^{(i)}$,
give preference to current particles most consistent with the next observation, $\by_t$, and where the
transition, $\tilde{P}$, moves particles to places that are compatible with $\by_t$. If the transitions $P(d\bx_t\mid \bx_{t-1})$
have densities with respect to $\tilde{P}(d\bx_t\mid \bx_{t-1})$, the
importance weights are well-defined, but the algorithm has complexity $O(N^2)$ because each 
unnormalised weight involves a summation over $N$ terms. 

The auxiliary particle filter of \cite{Pitt:1999} avoids this increase in
complexity by considering the target distribution on the product space of
the state at times $t-1$ and $t$ which is proportional to 
$$\sum_{i=1}^N w^i_{t-1}\delta_{\bx^i_{t-1}}(d\bx_{t-1}) 
P(d\bx_t\mid \bx_{t-1})g(\by_t\mid \bx_t) .$$
This is an approximation of $\pi_{t-1:t\mid t}$. If we use a proposal of the form
$$\sum_{i=1}^N \tilde{w}^i_{t-1} \delta_{\bx^i_{t-1}}(d\bx_{t-1}) 
\tilde{P}(d\bx_t\mid \bx_{t-1}),$$
the proposed pairs are obtained by first resampling the
particles $(\bx^i_{t-1})$ with probabilities $(\tilde{w}^i_{t-1})$
and then propagating the resampled particles with the transition
$\tilde{P}$. If we draw the sample
$(\bx^{A(i)}_{t-1},\bx^i_{t})$ from this proposal, then its importance weight is
\begin{equation}
  \label{eq:auxfilt}
  w^i_t \propto W^i_t = \frac{w^{A(i)}_{t-1}}{\tilde{w}^{A(i)}_{t-1}}
\frac{P(d\bx^i_{t}\mid \bx^{A(i)}_{t-1}) 
g(\by_t\mid \bx^i_{t})}{\tilde{P}(d\bx^i_{t}\mid \bx^{A(i)}_{t-1})}.
\end{equation}
In contrast to the bootstrap filter, the weights depend on the particles
at both times $t$ and $t-1$. The average of the un-normalised weights again
provides an approximation of $p(\by_t\mid \by_{1:t-1})$, and can be used to obtain
an unbiased estimate of the likelihood.


The second ratio in the equation for $W^i_t$ 
has to be understood as a Radon--Nikodym derivative.
If densities exist, it is simply the ratio of these densities. 
The auxiliary particle filter can be applied also to models where 
the observation distribution does not have a density because we 
observe part of the state without error. The formula for
the weights $W^i_t$ still makes sense provided we use a proposal density 
$\tilde{P}(d\bx_t\mid \bx_{t-1})$ that 
simulates states consistent with the new observation $\by_t$.

If $\tilde{w}^i_t=w^i_{t-1}$ and $\tilde{P}=P$, we
recover the bootstrap filter, but both the weights and the transition  
in the proposal can  depend on the new observation $\by_t$.
The optimal proposal in the sense of making the weights (\ref{eq:auxfilt})
constant is \cite[]{Doucet/Godsill/Andrieu:2000}
$$\tilde{w}^i_{t-1} \propto w^i_{t-1}p(\by_t\mid \bx^i_{t-1}), \quad 
\tilde{P}(d\bx_t\mid \bx_{t-1}) = P(d\bx_{t}\mid \bx_{t-1},\by_t).$$
These quantities are usually not tractable, but often one can
obtain good approximations with reasonable computing complexity. 

Even when the weights $w^i_{t}$ are constant, resampling will occur
at the beginning of the next iteration. The weights are then 
proportional to $p(\by_{t+1}\mid \bx^i_{t})$ whereas the bootstrap filter
has weights proportional to $g(\by_{t}\mid \bx^i_{t})=p(\by_{t}\mid \bx^i_{t})$. 
Since the former likelihood is flatter, 
the auxiliary particle filter has more equal weights, but the
difference is substantial only if the dependence in the state dynamics
is weak compared to the information from the observations. 

\subsection{Quasi-Monte Carlo Filters}

Quasi Monte Carlo methods achieve faster convergence rates than standard Monte Carlo
by replacing random draws by ``more regular'' samples. They start with
``low discrepancy'' points $(\bu^i)$ in the unit cube $[0,1)^d$, and transform
these into low discrepancy points from a general distribution of interest. The advantage of
quasi-Monte Carlo is that the error decays at a rate close to $1/N$, rather
than the $1/\sqrt{N}$ of standard Monte Carlo \cite[]{Niederreiter:1978}. Randomised versions of quasi Monte Carlo 
can even achieve rates close to $N^{-3/2}$ \cite[]{Owen:1998}. Though for high-dimensional
applications we need large $N$ to see any benefit from these quicker convergence rates \cite[]{Caflisch/Morokoff/Owen:1997}.

The first use of quasi Monte Carlo for particle filters was by
\cite{Fearnhead:2005JCGS}, though the computational cost was $O(N^2)$.
More recently, \cite{Gerber/Chopin:2015} have shown how quasi-Monte Carlo
can be applied
within a particle filter whilst still retaining the $O(N)$ computational
complexity. Their idea is to transform the state so that it is in
the $d$-dimensional unit cube and write the state transition as
$$\bX_t = \Gamma_t(\bX_{t-1}, \bU_t) \quad \bU_t \sim \U([0,1)^d)$$
where $\Gamma_t$ is an appropriate smooth function. Assuming that $(\bx^i_{t-1})$
is a quasi-Monte Carlo sample, one wants to modify the bootstrap
filter so that $(\bx^i_{t})$ is still a quasi-Monte Carlo sample.
For this, particles should not be resampled or propagated independently: if 
$\bx^i_{t-1}$ and $\bx^j_{t-1}$ are close, then the number of times they are resampled, $N^i_t+N^j_t$, should be as close as possible to $N(w^i_{t} + w^j_{t})$,
and they should be
spread out in the propagation step. 

In the one-dimensional case, $d=1$, this is  easy to achieve: we can
assume that the $x^i_{t-1}$ are in increasing order and that the
innovations $u^i_t$ are numbered such that the first $k$ points
$u^{1:k}_t$ have low discrepancy for any $k \leq N$. Then the
propagated particles 
$$x^i_{t} = \Gamma_t(x^i_{t-1},u^i_t) \quad (1 \leq i \leq N),$$
have the desired features. Similarly, for the balanced resampling
method discussed above, we arrange the subintervals in the same
order as the particles.

The difficulty in extending these ideas to higher dimensions is
how to define a suitable total order of points in the unit cube.
\cite{Gerber/Chopin:2015} use the Hilbert curve, which is a space-filling
fractal curve $H: [0,1) \mapsto [0,1)^d$ that preserves locality
and low discrepancy.

\subsection{Sequential Monte Carlo}

Particle filters have many applications outside of time series analysis.
They are then usually called sequential Monte Carlo algorithms and
produce samples from a complicated target distribution 
$\pi$ by recursive sampling from a sequence of $n$ intermediate
distributions $\pi_0, \pi_1, \ldots , \pi_n=\pi$. Here $\pi_0$ is
a distribution from which one can sample easily and we choose the sequence of
distributions so that any
two consecutive
distributions $\pi_{t-1}$ and $\pi_t$ are  ``close''. A prime
example is tempering, where
$$\pi(d\bx) \propto \phi(\bx) \pi_0(d\bx), \quad \pi_t(d\bx) 
\propto \phi(\bx)^{t/n} \pi_0(d\bx).$$

As in Monte Carlo filtering, sequential Monte Carlo produces a sequence
of particles by resampling, propagation with transitions $P_t$ and 
reweighting with weight functions $g_t$. If $(\bx^i_{t-1},
w^i_{t-1})$ is a weighted sample from $\pi_{t-1}$, then the propagated
particles $(\bx^i_{t}, w^i_{t-1})$ are a weighted sample from 
$$\pi_{t| t-1}(d\bx) = \int P_{t-1}(d\bx\mid \bx')
\pi_{t-1}(d\bx').$$
Therefore, the correct weight function $g_t$ is the density of 
$\pi_t$ with respect to $\pi_{t| t-1}$. In situations other than filtering,
this $g_t$ is intractable unless $P_t$ leaves $\pi_{t}$
invariant. Sequential Monte Carlo methods overcome this intractability
by working on the joint space of $(\bx_t,\bx_{t-1})$. They construct a joint
distribution whose marginal for $\bX_t$ is $\pi_t(\bx_t)$, and for which it
is then possible to calculate appropriate importance sampling weights. See 
\cite{DelMoral:2006} for more details.


\subsection{Ensemble Kalman Filter}
The ensemble Kalman filter has been developed in geophysics 
\cite[]{Evensen:94,Evensen:07}
and is used frequently in atmospheric physics, oceanography and
reservoir modeling. The propagation step is the same as in the
bootstrap  filter. For the update, the observations are assumed
to be linear combinations of the state with additive Gaussian noise: 
$\bY_t\mid  \bx_t  \sim \N(H\bx_t, R)$. If the prediction distribution
is normal, $\pi_{t| t-1} = \N(\bm_{t| t-1},P_{t| t-1})$, then the
filter distribution is also normal, $\pi_{t} = \N(\bm_{t},P_{t})$ with
$$\bm_{t}= \bm_{t | t-1} + K_t( \by_t - H \bm_{t | t-1}), \quad P_{t} = (I -
K_tH) P_{t | t-1}$$
where 
$$K_t = K(P_{t | t-1},R) = P_{t | t-1} H^T( HP_{t | t-1} H^T + R)^{-1}$$
is the Kalman gain. In the ensemble Kalman filter both $\pi_{t | t-1}$ and $\pi_t$ are approximated
by unweighted samples that we denote by $(\bx^i_t)$ and
$(\tilde{\bx}^i_t)$. The update step in the ensemble Kalman filter 
estimates $\bm_{t | t-1}$ and $P_{t | t-1}$ from the prediction sample
and then constructs the filter sample by transforming the prediction 
sample such that it has 
the mean and covariance given above. This can be done in different
ways.

The stochastic ensemble Kalman filter uses
\begin{equation}
  \label{eq:enkf-stoch}
  \tilde{\bx}^i_{t} = \bx^i_{t} + \hat{K}_t( \by_t - H \bx^i_t +
\eps^i_t), \quad \eps^i_t \sim \N(0,R).
\end{equation}
Square root filters use a deterministic affine transformation of
the sample $(\bx^i_{t})$. To define it, we introduce the matrix 
$\Delta \bx_{t}$ whose columns contain the centered particles
$\bx^i_{t}- \hat{\bm}_{t | t-1}$, and similarly $\Delta \tilde{\bx}_{t}$.
Then the mean is updated by
$$\hat{\bm}_{t} = \hat{\bm}_{t | t-1} + \hat{K}_t( \by_t - H \hat{\bm}_{t | t-1})$$
and the centered particles either by pre- or postmultiplication
$$\Delta \tilde{\bx}_{t} = A_t \cdot \Delta \bx_{t}, \quad \textrm{or } 
\Delta \tilde{\bx}_{t} = \Delta \bx_{t} \cdot W_t.$$
The matrices $A_t$ and $W_t$ are obtained by requiring that the 
sample $(\tilde{\bx}^i_{t | t})$ has the desired covariance
$$\Delta \tilde{\bx}_{t} (\Delta \tilde{\bx}_{t})^T = (N-1) \hat{P}_{t}=
(N-1) (I - \hat{K}_tH) \hat{P}_{t | t-1}.$$
This results in quadratic equations for $A_t$ and $W_t$ that can be
solved \cite[see][]{Tippett:03}. 
Postmultiplication is prefered for computational reasons if the sample size
$N$ is smaller than the dimension of the state.

Fig. \ref{Fig:enkf} shows an update step by the ensemble Kalman filter
in an example from numerical weather prediction.

\begin{figure}
 \centering
    \includegraphics[height=5cm]{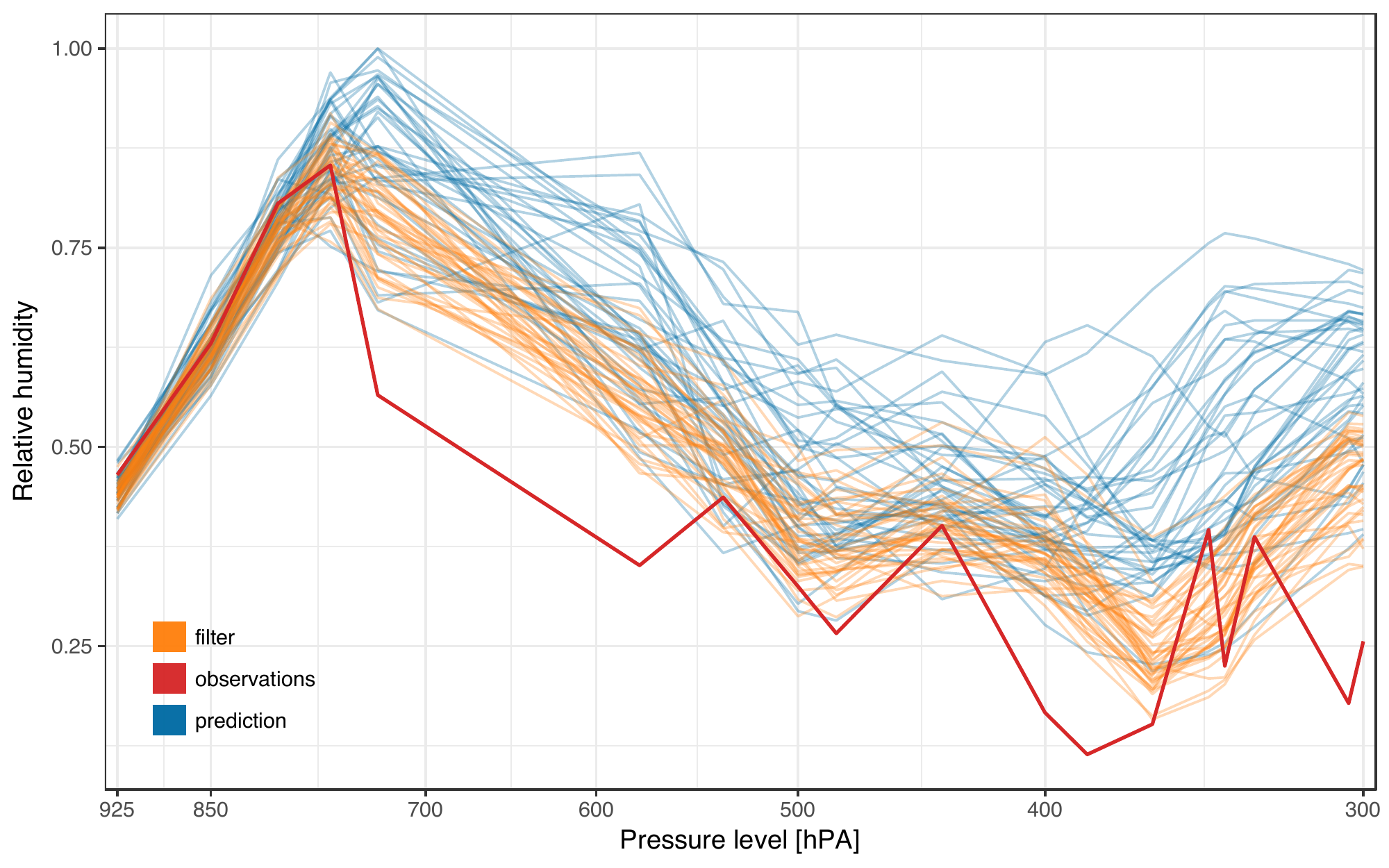} 
\caption{\label{Fig:enkf} 
Update of relative humidity as a function of pressure by the ensemble
Kalman filter using measurements
from one radiosonde launched on June 15, 2015 in Payerne, Switzerland.
The values of the state are averaged over different
pressure levels and a region near Payerne. The update moves the
prediction particles closer to the observations and reduces the spread.}
\end{figure}

For stability of the ensemble Kalman filter, the estimation of
the prediction covariance $P_{t | t-1}$ is crucial. We will come back
to this point briefly in Section \ref{Ssub:Stability}. There are various
methods to compute the update efficiently, depending on which version
is used and how $P_{t | t-1}$ is estimated, \cite[see e.g.][]{Evensen:03,Tippett:03}. If the state is high-dimensional,
$\hat{P}_{t | t-1}$ need not be computed or  stored, and the ensemble Kalman
filter has computational advantages over the population Kalman
filter, see \cite{Butala:09}.

The ensemble Kalman filter updates the particles by moving them
in space instead of weighting and resampling. It therefore does
not suffer from sample depletion like particle filters. It is
however biased in general, and can be viewed as reducing the variance
by allowing a bias. 


\subsection{Particle Filter Updates using Sequential Monte Carlo} 
If the observations are informative about the state, the two distributions
$\pi_{t | t-1}$ and $\pi_{t}$ are not close enough to make importance
sampling efficient. Using sequential Monte Carlo to go from $\pi_{t | t-1}$
to $\pi_t$ in a sequence of intermediate steps is therefore an attractive
idea. Intermediate steps can be defined either by tempering the likelihood,
$\pi_{t,\gamma} \propto \pi_{t | t-1}(d\bx_t) g(\by_t\mid \bx_t)^\gamma$, or --
if the dimension $d$ of $\by_t$ is large and the components of $\bY_t$
are independent given $\bx_t$ --
by the posterior given the first $k<d$ components of $\by_t$. However, in
order to obtain an algorithm that differs from a bootstrap filter,
the sequential Monte Carlo algorithm must include some propagation,
and the choice of the transition kernels used in propagating the particles
is difficult, in particular when the dynamics of the state is not
tractable and there is no analytic expression
for $\pi_{t | t-1}$. 

The simplest implementation of this idea is in \cite{Frei:2013} where
there is just one intermediate tempering of the likelihood and the
ensemble Kalman filter is used for  the first step while the particle filter is used
for the second. Both steps can be done analytically, and
sampling is only required for the next propagation step. For alternative 
approaches see \cite{Bunch:16} and \cite{Beskos:14}.


\subsection{Other Monte Carlo Filtering Algorithms}
Here we briefly discuss a few other filtering algorithms. Most of them try to reduce
the sample depletion problem of the particle filter and to improve
simultaneously on the ability of the ensemble Kalman filter in
non-Gaussian situations.

\subsubsection{Transport Filters}

\cite{Reich:13} proposes a linear deterministic update that replaces
resampling by averaging: instead of  
$\tilde{\bx}^i_{t} = \bx^{A(i)}_{t}$ he uses 
$$\tilde{\bx}^i_{t} = \sum_{j=1}^N \pr(A(i)=j) \bx^j_{t},$$
for some chosen resampling scheme that specifies the probabilities $\pr(A(i)=j)$ for all $i$ and $j$.
This approximation preserves the mean, but can be shown to have a reduced spread by
a simple application of  Jensen's inequality.

The approximation error depends on the chosen resampling scheme. 
The stochastic matrix $(p_{ij}) = (\pr(A(i)=j))$ must 
satisfy $\sum_i p_{ij} = N w^j_{t}$, but is otherwise arbitrary. 
\cite{Reich:13} chooses $(p_{ij})$ such that  on average the resampled particles are as close
as possible to the original particles. For a given distance matrix
$d_{ij}=d(\bx^i_{t},\bx^j_{t})$ between prediction particles, this means
we minimize 
$\sum_{i,j} d_{ij} p_{ij}$
subject to $p_{ij} \geq 0$, $\sum_j p_{ij}=1$ and $\sum_i p_{ij}=N
w^j_{t | t}$. This is a famous linear programming problem,
see e.g. \cite{Reich:15}, Section 7.4. If $(p_{ij})$ is the solution of this 
minimisation problem, the approximate update is then 
consistent as the sample size $N$ tends to infinity. Heuristically
this is true because most $p_{ij}=0$ in the optimal solution.

\subsubsection{Hybrid Filters} Hybrid filters combine particle and
ensemble Kalman filters with the goal of exploiting the advantages
of both methods. \cite{Reich:16} follow the idea of \cite{Frei:2013}, described above,
with a two-step update. They use the deterministic transport filter
of \cite{Reich:13} in the first step and the ensemble Kalman filter in the
second. 

\cite{vanLeeuwen:10} proposes a method to obtain equal weights by an
ensemble Kalman filter type proposal density. 

\subsubsection{Robust Filters} \cite{Calvet:15} propose a robust
filter to address the issue of observation outliers. They modify
the likelihood $g(\by_t\mid \bx_t)$ in order to reduce the impact of an
observation $\by_t$ that comes from a distribution other than the
nominal $g(\by_t\mid \bx_t)d\nu(\by_t)$. This has the additional benefit
of reducing sample depletion in the particle filter. 


\subsubsection{Rao--Blackwellisation}
For some models we can partition the state, $\bX_t=(\bX^{(1)}_t,\bX^{(2)}_t)$, such 
that $p(\bx_{1:t}^{(1)}\mid \bx_{1:t}^{(2)},\by_{1:t})$ is tractable. This most often
happens if, conditional on $\bx_{1:t}^{(2)}$, the model is linear-Gaussian. In this case
$p(\bx_{1:t}^{(1)}\mid \bx_{1:t}^{(2)},\by_{1:t})$ is Gaussian, with a mean and covariance that will
depend on $\bx_{1:t}^{(2)}$ but that can be calculated using the Kalman filter. In such cases we
can ``Rao--Blackwellise'' out the $\bX_t^{(1)}$ component of the state, and implement
a particle filter that targets $p(\bx_{1:t}^{(2)}\mid \by_{1:t})$. By reducing the dimension of the 
state-space, this latter particle filter can be much more efficient. See \cite{Doucet/Godsill/Andrieu:2000} and \cite{chen2000mixture} for examples
and further details.

\section{BASIC CONVERGENCE RESULTS} \label{S:Convergence}
There is now a substantial literature describing whether and how fast the particle filter
approximation converges to the filtering 
distribution as the number of
particles increases. Comprehensive results can be found in \cite{crisan2002survey}, 
\cite{del2004feynman} or \cite{Cappe:05}. \cite{chopin2004central} and
\cite{Kuensch:05} use  
a less general, but more direct approach. For  brevity, we limit ourselves
here to an intuitive discussion of some key ideas behind these convergence results. 

Let $\psi(\bx)$ be a suitable
test function, and let $\bX_t^i$ be the evenly-weighted particles 
of our particle filter at time $t$. Then the goal is to prove
an $L^p$-bound or a central limit theorem for the filter error
at time $t$,  
$$
\frac{1}{N}\sum_{i=1}^N \psi(\bX_t^i) - \int \psi(\bx_t)\pi_t(d\bx_t),
$$
as the number of particles, $N$, goes to infinity and the observations 
$\by_{1:t}$ are fixed. 

For fixed time $t$, such results hold under weak conditions on the
evolution and observation models, and typically one has the 
standard $1/\sqrt{N}$ error of a Monte Carlo procedure. But for
applications, one would like to know whether the  
bounds or the convergence are uniform in $t$. If the number
of particles required for a given accuracy of the estimated filter mean
needs to increase with the number of time steps, particle filters would be
of limited use. 

In addition to a sampling error at time $t$, the filter error has a second
component that occurs because the
particles $\bX_t^i$ are not sampled from the exact
filter distribution $\pi_t$, but from the distribution proportional
to $\sum_i P(d\bx_t\mid \bx^i_{t-1}) g(\by_t\mid \bx_t)$. The second component
of the filter error is thus the difference of the expectation of
$\psi(\bx_t)$ with respect to these two distributions. Both distributions
are obtained through a propagation step (\ref{eq:prop2}) and 
an update step (\ref{eq:update2}), applied to $\pi_{t-1}$ 
and $\hat{\pi}^N_{t-1}$ respectively. Therefore the second component
of the filter error is due to the error present at time $t-1$.
Uniform in time bounds require a control of error accumulation
as $t$ increases, or, equivalently, that the filtering distribution
$\pi_t$ forgets the initial distribution $\pi_0$ for $t \rightarrow
\infty$. 

Intuitvely, if the state process mixes well, then the error at 
time $t-1$ will be reduced when we go forward one time step using
a propagation and an update step. However, the update step can 
make this intuition invalid, and there are examples of state space models with
ergodic dynamics where the filter does not forget the initial distribution.
Forgetting of the initial distribution at a sufficiently fast rate does hold under 
an unrealistically strong condition of uniform mixing
of the state process. This assumption has been used in most
uniform-in-time convergence results for particle filters. Recently however,
\cite{Douc:14} have been able to prove such results under substantially
weaker conditions. See \cite{Atar:11} for a review of results about 
forgetting of the initial distribution by the filter.

By contrast, if there is strong dependence in the model, particle filters
can have poor Monte Carlo properties. This is most clearly seen when 
some components do not change at all in the state evolution, something
that occurs when we perform smoothing or when there are unknown fixed parameters -- see Sections \ref{S:Smoothing} and \ref{S:ParameterEstimation} respectively. In these cases the particle filter
variance will increase with $t$, and thus we need an increasing Monte Carlo
sample size as we analyse longer time-series. Often the Monte Carlo sample
size may need to increase exponentially with $t$.

\section{FILTER COLLAPSE} \label{S:Collapse}
\subsection{Importance Weights in High Dimensions}
In many examples, one observes that the maximal weight in the bootstrap
filter is very close to one, leading to the collapse of the 
filter. \cite{bengtsson2008curse}  provide theoretical insight into why this occurs 
by analysing cases where the dimension $d$ of the
observation $\by_t$ and the number of particles $N$ both go to infinity. 
Conditionally on $\by_t$, the log likelihood values $\log g(\by_t\mid \bx_t^i)$ 
then typically behave like 
a sample from $\N(\mu_d,\sigma_d^2)$ where $\sigma_d=O(\sqrt{d})$. 
If this holds, the ratio of the largest and the second-largest weight
can be approximated in distribution by
$$\exp(\sigma_d (Z_{(N)} - Z_{(N-1)}))$$
where $(Z_{(N-1)},Z_{(N)})$ are the two largest values of $N$
independent $\N(0,1)$-variables. By standard results from extreme
value theory (e.g. \cite{Embrechts:97}, Theorem 4.2.8), the difference
$Z_{(N)}-Z_{(N-1)}$ is of the order $O_P(1/\sqrt{2 \log N}$). Therefore
the maximal weight converges to one in probability unless $N$ grows
exponentially with $d$. For the two largest
weights to be asymptotically equal, we even need $\log(N)/d \rightarrow \infty$.

For related results on the required Monte Carlo sample size for importance sampling see
\cite{chatterjee2015sample}, \cite{agapiou2015importance} and 
\cite{Sanz-Alonso:16}.


\subsection{Stability of the Ensemble Kalman Filter}
\label{Ssub:Stability}
\cite{LeGland:11} and \cite{Frei:13} have studied the asymptotics
of ensemble Kalman filters in the standard setting where we fix the
dimension of the states and observations and increase the number of particles, 
$N\rightarrow\infty$. For practical applications, the more relevant question is
whether the filter does not lose track of the state when 
$N$ is smaller than the  dimension of the state. Some modifications of the method
are necessary to achieve this stability, and we still lack a full
understanding of the problem. 

One reason for instability is sampling errors in the estimated
prediction covariance $\hat{P}_{t | t-1}$. Particularly harmful are
underestimation of the diagonal elements, leading to overconfidence
in the prediction sample, and spurious non-zero off-diagonal elements
in cases with sparse observations. This is because the ensemble Kalman
filter works by first updating the observed components and then
regressing unobserved components on these updates. Techniques to
mitigate these problems are inflation of the diagonal elements and regularization
of estimated covariances by tapering, that is elementwise multiplication
of $\hat{P}_{t | t-1}$ with a band-limited correlation matrix. These techniques
are effective, but the choice of the tuning constants is difficult. 
A method which is similar, but not equivalent, to covariance tapering
is localization. Since localization is used also for particle filters,
we discuss it separately in the next subsection.

\cite{Kelly:15} present a rigorous analysis of how the unmodified 
ensemble Kalman filter can diverge to infinity even when the dynamics of
the state is deterministic with a stable fixed point. \cite{Tong:16}
propose an adaptive inflation of the diagonal of $\hat{P}_{t | t-1}$
such that the Markov process consisting of the state $\bX_t$ and
the filter ensemble $(\bX^i_t)$ is geometrically ergodic provided 
the state dynamics has a Lyapunov function. Hence the ensemble cannot
diverge to infinity, but there is no information about how close
the state $\bX_t$ and the filter mean are. 

\subsection{Preventing Collapse by Localization}
In many applications with high-dimensional states or observations, the 
components of $\bX_t$ and $\bY_t$ are associated with positions in space.
Usually in such cases the observation distribution is local, 
$$g(\by_t\mid \bx_t) = \prod_v g(y_{t,v}\mid x_{t,N(v)}),$$
where $v$ is used to indicate components of $\by_t$ and $N(v)$ is a
set of components of $\bx_t$ whose positions are close to $v$. If dependence in 
the predictive distribution $\pi_{t | t-1}$ is small between regions
far apart, then intuitively it seems reasonable to use local updates
where any given component of $\bx_t$ is only affected by close-by components
of $\by_t$. However, the exact update is typically not local
as can be seen in the example where under $\pi_{t | t-1}$ the state is a circular
Gaussian moving average of order 1 and $\bY_t \mid  \bx_t \sim \N(\bx_t,I)$.
Still, a local update should be close to optimal and more stable because it 
combines updates in much lower dimensions. In addition, local updates can
take advantage of modern, highly parallel, computer architectures.

For the ensemble Kalman filter update, localization was 
introduced early on \cite[see e.g.][]{Evensen:03,Ott:04}, 
and is now a well-established 
technique. It can be achieved by
imposing sparsity on the estimated Kalman gain, $\widehat{K}_t$. 
Instead of simply setting most entries of the gain equal to zero,
updates with better smoothness properties can be obtained by artificially
increasing the observation error variance to infinity as the distance to the
position to be updated increases. See \cite{Hunt:07} for details and further
discussion. 

The animations in the supplemental material illustrate covariance
tapering and localization in the Lorenz 96 model.

For particle filters or hybrid methods, localization is much more difficult
because any kind of resampling that does not occur globally
for all components introduces artificial discontinuities in the 
particles. Such discontinuities can have drastic consequences in
the next propagation step if the transition depends on differences
between neigboring components which is typical in many applications. 
\cite{Robert:17} have proposed a method that introduces a smooth transition
between regions with different resampling by making partial Gaussian
assumptions for $\pi_t$. It is similar but not identical to the idea
in \cite{Bengtsson:03}. \cite{Rebeschini:15} analyze theoretical properties
of a particle filter which partitions the set of positions into blocks
and applies independent resampling for state variables in each block.
Their error bounds are independent of dimension and
small at positions away from boundaries between blocks. It would
be interesting to extend these results to methods that also reduce
the discontinuities between blocks.

\section{PARTICLE SMOOTHING} \label{S:Smoothing}


We now turn to the related problem of smoothing. This involves calculating, or approximating, 
the distribution of past values of the state. We will consider three related smoothing problems. 
The first is full smoothing, calculating the conditional distribution of the complete
trajectory of the state, $\pi_{0:t\mid t}(d\bx_{0:t})$. The others are fixed-lag smoothing, where
our interest is only in the trajectory of the state at the most recent $L+1$ time-points, $\pi_{t-L:t\mid t}(d\bx_{t-L:t})$; and
marginal smoothing, where our interest concerns the state at a fixed time-point $\pi_{s | t}(d\bx_{s})$ for some $s<t$. If we can
solve the full smoothing problem well, then this immediately gives us a solution to the fixed-lag and marginal smoothers. However,
as we will see, there are approaches that can work well for the latter two problems but not the former.

The auxiliary particle filter of Section \ref{S:AuxiliaryParticleFilter} gives a solution to the fixed-lag smoothing problem for a lag of $L=1$. When calculating the filtering
distribution at time $t$ it generates weighted particles that consist of states at both time $t$ and $t-1$. These weighted particles approximate the joint density $\pi_{t-1:t\mid t}(d\bx_{t-1:t})$.
This idea has been extended by \cite{Doucet/Briers/Senecal:2006}, though for larger $L$ this approach suffers from the need to define an efficient proposal distribution for $\bx_{t-L+1:t}$, which can be 
difficult.

\cite{Kitagawa:1996} noted that we can trivially adapt a particle filter to solve the full smoothing problem by just storing the
trajectory associated with each particle. Thus at time $t-1$ a particle will consist of a realisation of the full state trajectory 
$\bx_{0:t-1}$. When we perform the propagation step of the particle filter at time $t$ the new particle will be the concatenation of its
ancestor particle at time $t-1$ and the simulated value for the state at time $t$. 
The resulting algorithm can be viewed as a particle filter approximation to
recursions (\ref{eq:prop1}) and (\ref{eq:update1}) rather than to
(\ref{eq:prop2}) and (\ref{eq:update2}).  
It incurs an additional storage cost over the basic particle filter, but otherwise shares the same computational properties. 

However, the algorithm is impracticable in most applications. 
When we re-define our particle filter so that its particles are the full trajectory of the state, 
the value of $\bx_{0:s}$ of any particle at time $t>s$ will necessarily be equal to one of the particles from time $s$. Furthermore the number of distinct paths for $\bx_{0:s}$ will decrease monotonically as
$t$ increases. If $t-s$ is sufficiently large, all particles will share the same value of $\bx_{0:s}$; \cite{jacob2015path} show that this will almost surely happen for a time $t$ such that $t-s=O(N\log N)$. 

Thus we will observe particle degeneracy in earlier parts of the particle trajectories. This can be seen, for the stochastic volatility model, from the left-hand column of plots in Figure \ref{Fig:Smoother}. In each
case the particle approximation for the smoothing distribution of $\bX_s$ given $\by_{1:t}$ degenerates to a single distinct value for $s\ll t$. 

\begin{figure}
\centering
\begin{subfigure}[b]{0.475\textwidth}
 \centering
   \includegraphics[width=0.9\textwidth]{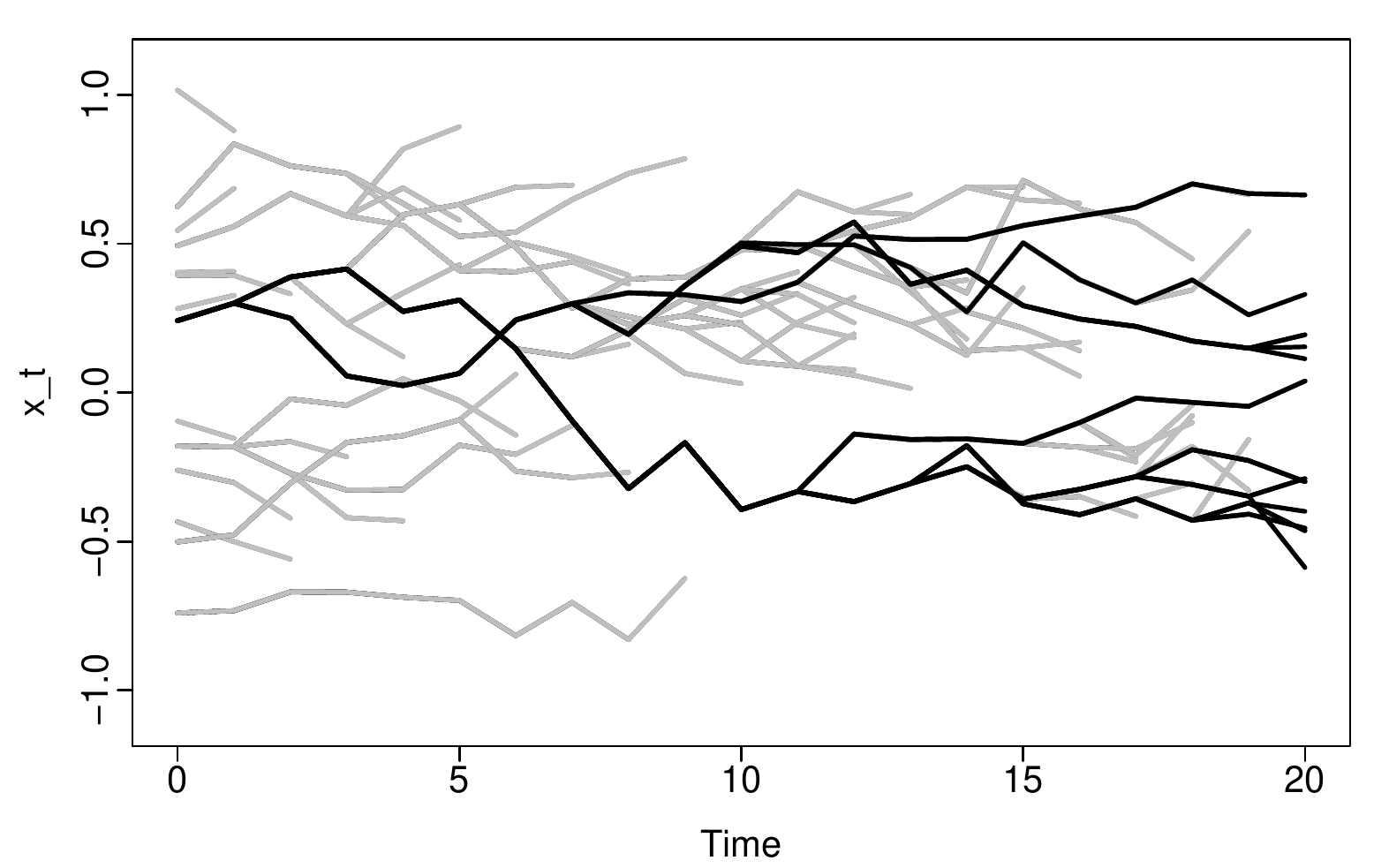}   
\end{subfigure}
\begin{subfigure}[b]{0.475\textwidth}
 \centering
   \includegraphics[width=0.9\textwidth]{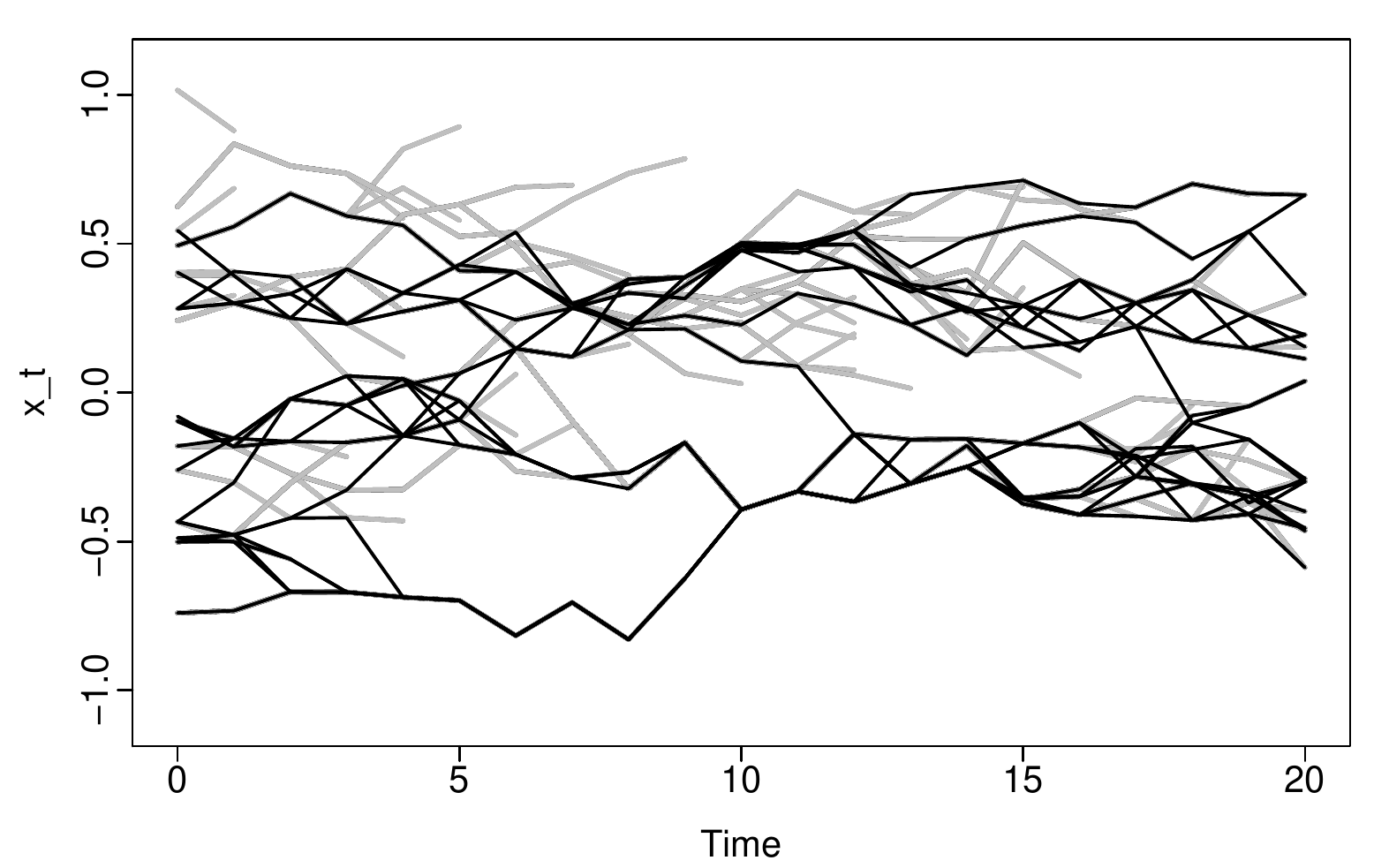}
\end{subfigure}
\vskip\baselineskip
\begin{subfigure}[b]{0.475\textwidth}
 \centering
   \includegraphics[width=0.9\textwidth]{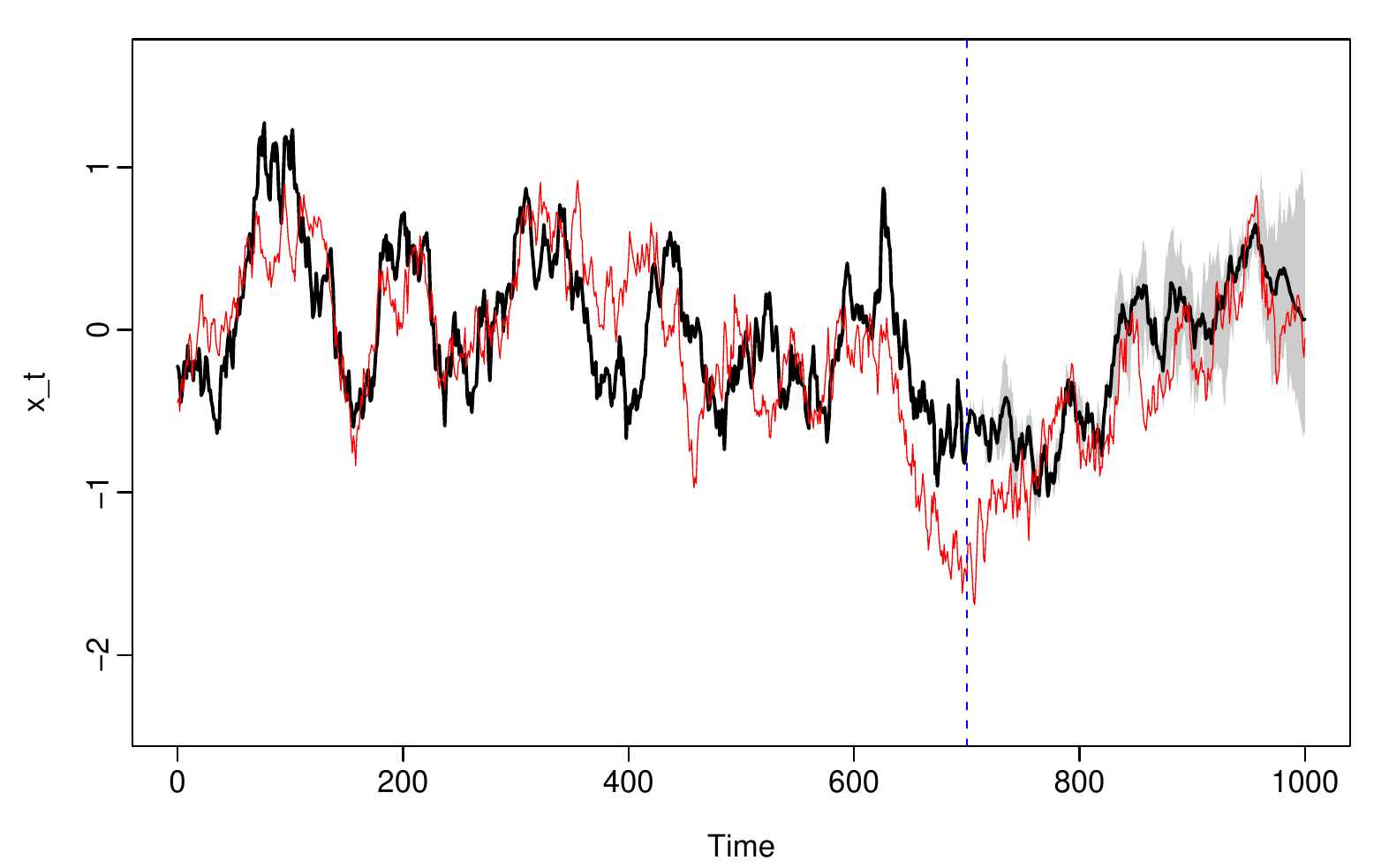}
\end{subfigure}
\begin{subfigure}[b]{0.475\textwidth}
 \centering
   \includegraphics[width=0.9\textwidth]{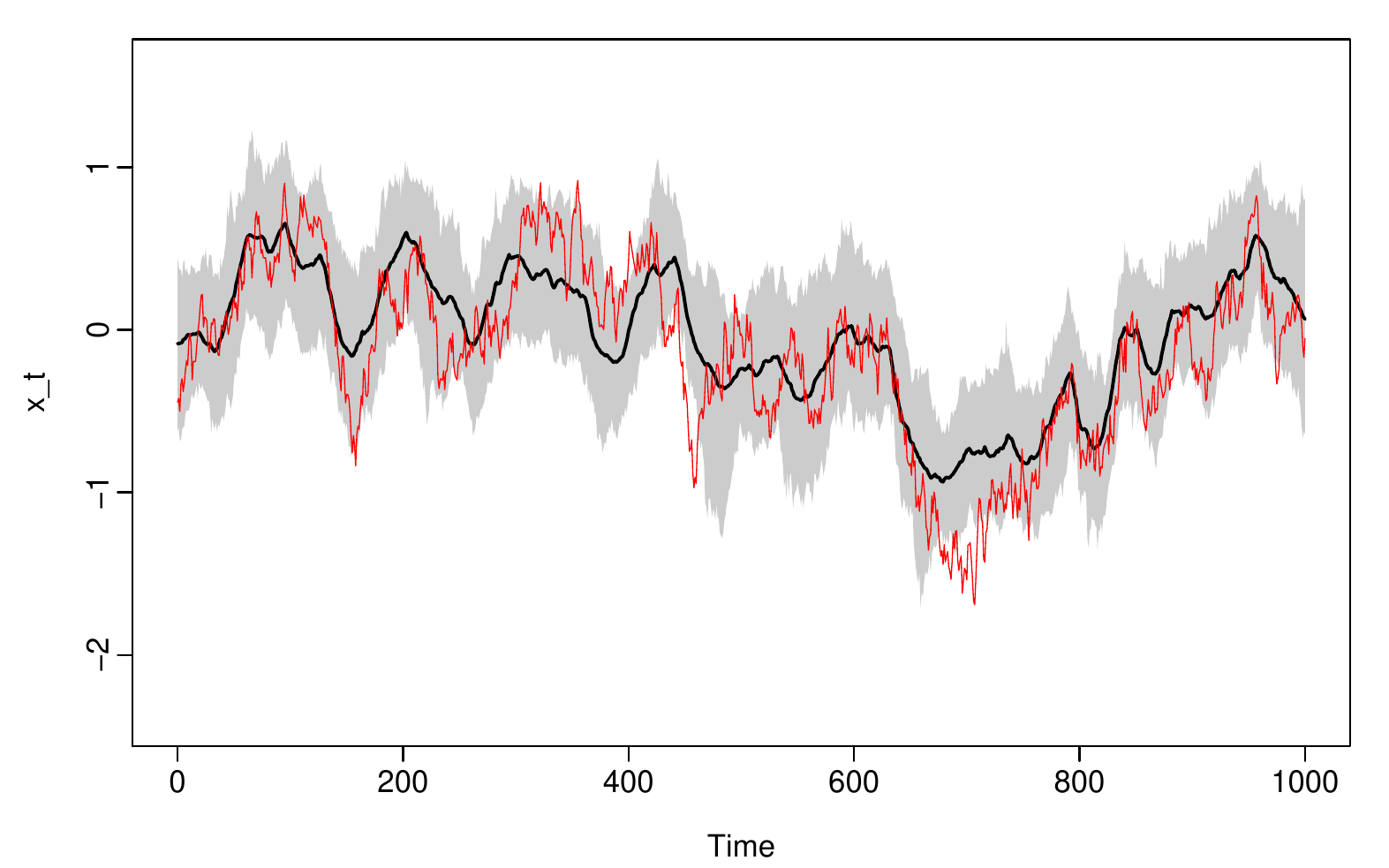}
 \end{subfigure}      
\caption{\label{Fig:Smoother} 
Left-hand column: results from the simple smoother of \cite{Kitagawa:1996} for the stochastic volatility model; right-hand column: results from the forward-backward smoother. Top  plots show the paths stored after each
iteration of the  \cite{Kitagawa:1996} smoother, in grey, and the final
sample of paths, in black. For the simple smoother the latter coincide with
all particles at the final time-point, but their diversity is reduced as
we go back towards the start. The forward-backward smoother is able to sample new paths during the backward simulation step, and thus maintains sample diversity throughout. Bottom plots show estimated
mean and 95\% coverage intervals for the state given data up to time 1000. For each plot the mean is in black, coverage intervals are in grey and the true state in red. The blue dashed vertical line shows the point
at which all particles at time $1,000$ have converged to single value for
the state. We used 20 and 200 particles for the top and bottom
plots respectively. In the supplemental material, the top-left plot is
available as an animation.}
\end{figure}

Whilst this simple algorithm of \cite{Kitagawa:1996} is in practice not
suitable for the full smoothing problem, in some situations it can give
good results for fixed-lag smoothing. There is some indication of this from the bottom-left plot of Figure  \ref{Fig:Smoother}, with the smoothed coverage intervals for $\bX_s$ demonstrating reasonable particle diversity
for the most recent time-points. 

Furthermore, the simple smoother can be used to approximately solve the marginal smoothing problem if we are willing to assume that observations sufficiently far in the future have little information about the current
state. This motivates an approximation whereby, when estimating the state at time $s$, we ignore any observations after a time $s+L$ for some suitably chosen lag $L$. Mathematically, this equates to assuming
$\pi_{s | n}(d\bx_s)\approx \pi_{s | s+L}(d\bx_s)$ for $n>s+L$ and large enough $L$ \cite[see e.g.][]{polson2008practical}. The algorithm
of \cite{Kitagawa:1996} can be used to approximate $\pi_{s | s+L}(d\bx_s)$ for a suitable value $L$, and then this approximation is used as an approximation to $\pi_{s | n}(d\bx_s)$ for subsequent times $n>s+L$. Whilst
the assumption underlying this approach is often reasonable, choosing an appropriate $L$ can be difficult in practice.

\subsection{Forward-Backward Particle Smoother}

Improvements on the simple smoother of \cite{Kitagawa:1996} are possible. One approach involves an additional, backwards smoothing recursion to post-process the output of the particle filter.
For this method we require that the state-transition distribution
has a density, denoted by $p(\bx_{s+1}\mid \bx_s)$, and that this 
density is analytically tractable. If state-transition densities
exist, then also all prediction and filter distributions have densities,
denoted by $\pi_s(\bx_s)$ and $\pi_{s | s-1}(\bx_s)$.

The forward-backward particle smoother is based on the same ideas that are used
within the forward-backward algorithm for discrete-state hidden Markov
models and linear Gaussian state space models \cite[]{Baum:1970,Durbin:01}. It
applies to all smoothing problems, and we begin by describing how
it is used to simulate from an approximation to the full smoothing
distribution \cite[]{Godsill/Doucet/West:2004}. 

By the conditional independence of $(\bx_{s+2:t},\by_{s+1:t})$ and
$(\bx_s,\by_{1:s})$ given $\bx_{s+1}$ and Bayes' formula
\begin{equation}
\label{eq:backtrans}
p(\bx_s\mid \bx_{s+1:t},\by_{1:t})=p(\bx_s\mid \bx_{s+1},\by_{1:s}) =
\frac{p(\bx_{s+1}\mid \bx_s)\pi_{s}(\bx_s)}{\pi_{s+1\mid s}(\bx_{s+1})}.
\end{equation}
This shows that under the smoothing distribution the state is still
a Markov process with backward transitions that are proportional to
the marginal forward transitions times the filter densities. 
Thus given a particle approximation to $\pi_{s}$, $\{\bx_{s}^{i},w_{s}^{i}\}$, 
we can construct a particle approximation to $p(\bx_s\mid \bx_{s+1},\by_{1:s})$.
This will have the same particles as the
the filter approximation, but with modified weights that are proportional 
$w_s^i p(\bx_{s+1}\mid \bx_s^i)$. If we have run the particle filter
forward in time, and stored the particle approximations to all filtering
distributions, we then have a backward simulation step: 
\begin{itemize}
 \item[(1)] Simulate $\bx_{t | t}$ from the particle approximation to
 $\pi_t(d \bx_t)$.
\item[(2)] For $s=t-1,\ldots,0$, simulate $\bx_{s | t}$ conditional on 
$\bx_{s+1\mid t}$ from the discrete distribution which assigns 
probability proportional to
$ w_s^{i} p(\bx_{s+1\mid t}\mid \bx_s^{i})$ to value $\bx_s^{i}$.
\end{itemize}
The cost of simulating one realisation of the trajectory is $O(tN)$,
and the cost of simulating a sample of $N$ trajectories is $O(tN^2)$.

To obtain a weighted approximation of the marginal smoothing density,
we integrate out the state at time $s+1$:
\begin{equation} \label{eq:FBsmoother}
  \pi_{s | t}(\bx_s)=\pi_{s}(\bx_s)\int \frac{p(\bx_{s+1}\mid \bx_s)}
{\pi_{s+1\mid s}(\bx_{s+1})}\pi_{s+1\mid t}(\bx_{s+1})d \bx_{s+1}. 
\end{equation}
Thus given a particle approximation to $\pi_{s}$, $\{\bx_{s}^{i},w_{s}^{i}
\}$, and one to $\pi_{s+1\mid t}$, $\{\bx_{s+1\mid t}^{i},w_{s+1\mid t}^{i} \}$,  
we can construct a particle approximation to $\pi_{s | t}$ \cite[]{Hurzeler/Kunsch:1998,Godsill/Doucet/West:2004}. As in the
above algorithm, it has the same 
particles as the filter approximation, but with new weights 
\[
w_{s | t}^{i}=w_{s}^{i}\sum_{j=1}^N w_{s+1\mid t}^{j} \left( 
\frac{p(\bx_{s+1\mid t}^{j}\mid \bx_{s}^{i})}
{\sum_{k=1}^N  w_{s}^{k} p(\bx_{s+1\mid t}^{j}\mid \bx_{s}^{k})}
\right) \mbox{ for $i=1,\ldots,N$.}
\]
The calculation of smoothing weights at time $s$ involves considering all
pairs of particles at times 
$s$ and $s+1$, and hence has an $O(N^2)$ cost. If we wish to calculate the
smoothing distribution at time $t-L$, then the cost of the smoothing
iterations will be $O(LN^2)$. We get for free  
all smoothing distributions from time $t$ to time $t-L$. 

We can see the improvement that the forward-backward smoother gives by comparing the plots from the right-hand column of Figures \ref{Fig:Smoother} to those of the left-hand column. 
In particular these show how the forward-backward smoother is able to maintain particle diversity, and thus give a reasonable approximation to the smoothing distribution, at all time-points.

A closely related particle smoothing algorithm is the two-filter smoother \cite[]{Kitagawa:1996}. This involves running two independent particle filters, one forward in time and one backward in time. The output of these
filters at any time point $s$ can then be combined to obtain a particle approximation to the smoothing distribution. See \cite{Briers/Doucet/Maskell:2010} for more detail. 
Also see \cite{Hurzeler/Kunsch:1998}, \cite{Fearnhead/Wyncoll/Tawn:2010} and \cite{douc2011sequential} for importance sampling and 
rejection sampling approaches to reduce the complexity of either the forward-backward smoother or the two-filter smoother to linear, rather than quadratic, in the number of particles.

\subsection{Ensemble Kalman smoothing}

The trivial adaptation of the particle filter which uses the trajectory
$\bx_{0:t}$ instead of the value $\bx_t$ at time $t$ works also for
Ensemble Kalman filter, see \cite{Evensen:03}, Appendix D.
The stochastic version uses, in addition to (\ref{eq:enkf-stoch}), 
the following update for the particles $\bx^i_{s | t}$ at times $s<t$:
$$\bx^i_{s | t} = \bx^i_{s | t-1} + \widehat{K}_{s,t}(\by_t - H \bx^i_{t} + \eps^i_t),$$
where the cross-gain $\widehat{K}_{s,t}$ is based on an estimate of the
cross-covariance between $(\bx^i_{s | t})$ and $(\bx^i_t)$. 
Weak points of this method are the restriction to linear dependence
between $\bX_s$ and $\bX_t$ and the need to estimate many cross-covariances.

There is also an analogue of the forward-backward marginal particle
smoother, see \cite{Stroud:10}. In a linear Gaussian model, the recursion
(\ref{eq:FBsmoother}) allows  us to express the first and second moments of $\pi_{s | t}$
in terms of the first and second moments of $\pi_{s+1\mid t}$ and $\pi_{s}$. Using
this one can derive an approximate sample $(\bx^j_{s | t})$ by a transformation of
the samples $(\bx^j_{s})$ and $(\bx^j_{s+1\mid t})$.

In numerical weather prediction, so-called
four-dimensional variational data assimilation is used frequently.
It computes the posterior mode of $\pi_{t-L:t}$ numerically, but lacks
uncertainty quantification. Hybrid methods which combine variational
data assimilation with ensemble Kalman filters have also been developed,
see \cite{Bannister:17}.


\section{PARAMETER ESTIMATION} \label{S:ParameterEstimation}

So far we have ignored the issue of parameter estimation within the filtering and smoothing problems. The algorithms we have presented have been suitable for inference conditional on knowing the parameter values 
of the underlying model. We now turn to problems where the parameters are unknown. We will denote the vector of parameters by $\btheta$, and write $P_{\btheta}$ and $g_{\btheta}$ for the transition distribution of the state
model and the likelihood function respectively. We can still apply the filtering recursions, for example (\ref{eq:prop2}) and (\ref{eq:update2}), but these will be conditional on a parameter value. 
As such
we will write, for example, 
$\pi_t(d\bx_t\mid \btheta)$ to be the filtering distribution conditional on given parameter value $\btheta$.

We will consider two different situations where we wish to estimate parameters. The first is on-line parameter estimation and the second is batch estimation. For the latter we only consider a relatively recent class
of algorithms, particle MCMC, which embed a particle filter within an MCMC algorithm. Our overview of on-line parameter estimation methods is deliberately brief, and the reader is referred to \cite{Kantas:2015} for
a recent article length review of both on-line approaches and related
approaches to batch estimation. \cite{Kantas:2015} discuss particle MCMC
methods only briefly, hence our stronger focus on those methods here.

\subsection{Online Parameter Estimation} \label{S:OnlineParameter}

Issues and methods for online parameter estimation within a particle filter are closely linked to those for particle smoothing. This stems from the fact that many quantities, such as the likelihood or score function, or
even the posterior distribution for the parameters, can be written as expectations with respect to the smoothing distribution. The first link between the two problems is that, just as for smoothing, there is 
a trivial extension to a particle filter that can deal directly with estimating the parameters, but that is rarely useful in practice.

This extension applies when we have a prior distribution for the parameter. In this case we can extend the state of our model to incorporate the parameter vector. So we have a new state, 
$\bx'_t=(\bx_t,\btheta)$ say. We can trivially write down the state evolution and observation model for $\bx'_t$, and apply a particle filter to approximate the filtering distribution
$\pi_t(d\bx'_t)$.  However the dynamics of
this particle filter will 
leave the parameter component unchanged at each
iteration. As a consequence of this deterministic update,
the number of distinct parameter values of the particles can only decrease
at each iteration, 
and often the filter's approximation to the posterior for $\btheta$
reduces very quickly to having a handful, or even just one, distinct
particle value.

There have been three main approaches to overcome the particle degeneracy that necessarily occurs in this simple method:

The simplest way to avoid this degeneracy is to break the ties by 
adding a small random noise to the parameter component of the particles
after resampling.
This effectively means that we use a kernel density
approximation of the filter distribution instead of a mixture of point masses.
The most effective version of this
idea is that of \cite{Liu/West:2001}, which shrinks the filter particles 
towards their mean before adding the noise.  
This ensures that the kernel density approximation has both the same mean
and variance as the 
original particle approximation. Without this shrinkage the approximation
of the kernel density estimation leads to an increase in the variance of
our approximation of the posterior  
at each iteration. These can accummulate and lead to substantial over-estimation of the parameter uncertainty. The algorithm of \cite{Liu/West:2001} has been shown to perform well in some applications, but it
lacks any theoretical guarantees, and has tuning parameters, such as the kernel bandwidth, that can be hard to choose.

An alternative approach is to use MCMC moves within the particle filter to sample new parameter values for each particle using an MCMC kernel that has the current posterior
distribution as its invariant distribution. The most common choice of MCMC kernel is a Gibbs kernel, that samples a new $\btheta$ for a particle from the parameter's conditional distribution given
the particle's stored trajectory. In many situations this distribution depends on the trajectory through some low-dimensional summary statistics, and we need only store and update these summaries, as opposed
to storing the full trajectory. The initial idea of using MCMC with a particle filter comes from \cite{Fearnhead:1998}, though the original use of such MCMC moves to update parameters was in \cite{Gilks/Berzuini:2001}, and
the use of summary statistics was suggested by \cite{Storvik:2002} and \cite{Fearnhead:2002}. Recently the general idea of using MCMC to update parameter values for models where sufficient statistics exist has been termed particle learning
\cite[]{Carvalho:2010}. Whilst using MCMC steps within the filter does reduce the problem of degeneracy within the particle filter, it does not remove it, because the updates for the parameters depend on summaries of the trajectory of the state. As mentioned above, the particle filter's approximation to the smoothing distribution of the trajectory will also
degenerate \cite[see Section 7.2 of][for a thorough, empirical evaluation of this method]{Kantas:2015}.

The third approach is to use some form of stochastic approximation method. The idea is to have a current estimate of the parameter at each iteration. The particle filter update at iteration $t$ is performed conditional
on the current parameter estimate, $\hat{\btheta}_t$. Simultaneously the particle filter is used to estimate the score function, that is the gradient of the log-likelihood, at $\hat{\btheta}_t$ and this
gradient information is used to update the estimate of the parameter. See \cite{Poyiadjis:2011}, \cite{nemeth2016particle} and \cite{olsson2017} for further details.

\subsection{Particle MCMC} \label{S:PMCMC}

We now consider batch estimation of parameters. That is, we assume we have been given a fixed data set $\by_{1:n}$, from which we wish to estimate parameters of our model. We are no longer constrained to methods
that are sequential or online, though methods for online parameter estimation, together with simple extensions of them, can still be applied 
\cite[see][]{Poyiadjis:2011,Kantas:2015}.

We will focus on one specific class of methods, particle MCMC \cite[]{Andrieu/Doucet/Holenstein:2010}. These are MCMC methods that target the joint posterior of the parameter and the state, and that
use particle filter methods to develop novel, and hopefully efficient, proposal distributions. The most basic particle MCMC algorithm is a form of pseudo-marginal MCMC algorithm \cite[]{Andrieu/Roberts:2009}
that leverages the fact that a particle filter gives an unbiased estimate of the likelihood -- see Section \ref{S:BootstrapFilter}. However a particle filter approximation to a Gibbs sampler has also been developed.
We describe these two approaches in turn. While particle MCMC was initially derived based on using particle filters to improve MCMC algorithms, it is also possible to embed particle MCMC methods
within particle filters \cite[]{Chopin:2013,Fulop/Li:2013}.

\subsubsection{Particle Metropolis Hastings}

Assume we have a prior density $p(\btheta)$ for our parameter vector. The
posterior density $p(\btheta\mid \by_{1:n})$ is then proportional to 
$p(\btheta)L(\btheta)$ where $L(\btheta)=p(\by_{1:n}\mid \btheta)$ is the likelihood.
If we run a particle filter conditional on parameter $\btheta$, we can obtain an unbiased estimate $\hat{L}(\btheta)$
of $L(\btheta)$. The particle marginal Metropolis-Hastings algorithm simulates a Markov chain with state, $(\btheta,\hat{L}(\btheta))$, which consists of the parameter vector and an estimate of the likelihood. Given a
proposal distribution for the parameter, with density $q(\btheta'\mid \btheta)$, the algorithm iterates the following steps
\begin{itemize}
  \item[(0)] Assume the current state at iteration $i$ is $(\btheta_i,\hat{L}_i)$, where $\hat{L}_i$ is an unbiased estimate of $L(\btheta_i)$.
 \item[(1)] Propose a new parameter vector $\btheta'\sim q(\btheta'\mid \btheta_i)$.
 \item[(2)] Run a particle filter, conditional on parameter vector $\btheta'$ to get $\hat{L}'$, an unbiased estimate of $L(\btheta')$.
 \item[(3)] With probability
 \[
  \min \left\{ 
  1, \frac{p(\btheta')q(\btheta_i\mid \btheta')\hat{L}'}{p(\btheta)q(\btheta'\mid \btheta_i)\hat{L}_i}
  \right\}
 \]
 set $(\btheta_{i+1},\hat{L}_{i+1})=(\btheta',\hat{L}')$, otherwise $(\btheta_{i+1},\hat{L}_{i+1})=(\btheta_i,\hat{L}_i)$.
\end{itemize}
The acceptance probability in step (3) is just the standard Metropolis-Hastings acceptance probability, except the true likelihood values are replaced by their unbiased estimates. Perhaps surprisingly, the
resulting algorithm still has the posterior for $\btheta$ as the $\btheta$-marginal of its stationary distribution. To see this, denote by $\bU$ the set of random 
variables used within the particle filter, and let  $L(\bU,\btheta)$ be the estimator of the likelihood we get from the particle filter run using random variables, 
$\bU$, with parameter $\btheta$. Then the particle marginal Metropolis-Hastings algorithm is a standard Metropolis-Hasting algorithm with target density proportional to
$p(\btheta)\pr(d\bu)L(\bu,\btheta)$ and with proposal density
$q(\btheta'\mid \btheta)\pr(d\bu)$, but that stores $(\btheta,L(\bu,\btheta))$ rather than $(\btheta,\bu)$. The marginal density
of this target is the posterior density for $\btheta$ as
\[
 \int p(\btheta)\pr(d\bu)L(\bu,\btheta) = p(\btheta)
\E \left(L(\bU,\btheta) \right)=p(\btheta)L(\btheta),
\]
by the unbiasedness of the estimator of the likelihood from the particle filter. 

The quality of the approximation to the likelihood affects how well the resulting algorithm mixes
\cite[]{Andrieu/Vihola:2015,Andrieu/Vihola:2016}. The benefit of particle MCMC is that, for well-mixing models, there is evidence that as the number of observations increases we only
need the number of particles used to increase linearly to maintain a similar level of mixing of the MCMC algorithm. This results in a computational complexity that is quadratic in the number of observations.

It is straightforward to extend the above particle marginal Metropolis-Hastings algorithm so as to obtain samples from the joint posterior for the parameter and the state, $\bx_{0:n}$. In step (2) we run a particle
filter that stores the state trajectory for each particle and outputs both our unbiased likelihood estimate, $\hat{L}'$, and a sample trajectory. We then accept or reject the new parameter value, the unbiased estimate 
and the trajectory in step (3). For further details see \cite{Andrieu/Doucet/Holenstein:2010}.

There is flexibility within the above particle MCMC algorithm, in terms of the choice of proposal distribution for the parameter, the number of particles to use in the particle filter algorithm, and the version of
particle filter used.  As we get better estimates of the likelihood, we may expect the particle MCMC algorithm to behave increasingly like an MCMC algorithm using the
true likelihood values, so our choice for proposal distribution can be informed by experience from implementing standard MCMC algorithms. Better proposal distributions for the underlying exact MCMC algorithm
should lead to better mixing of the particle MCMC algorithm. In particular
this has led to particle versions of Langevin algorithms that leverage the particle filter's ability to estimate gradient information so
as to improve the proposal distribution \cite[]{Dahlin:2014,Nemeth:2016}.

Theory also shows that the better the estimate of the likelihood, the more efficient the particle MCMC algorithm will be. This suggests using the most efficient particle filter algorithm available for a given problem.
It also shows that increasing the number of particles will improve mixing. However, this comes with an increased computational cost of running the filter. There have now been a number of theoretical studies linked
to choosing the optimal number of particles, so as to trade-off better mixing with the increased computational cost. The first results for this were from \cite{Pitt:2012}, and these have been extended by
\cite{Sherlock:2015}, \cite{Doucet/Pitt/Deligiannidis/Kohn:2015} and \cite{Nemeth:2016}. The main conclusion is that we should tune the number of particles so that the variance 
of our estimate of the log-likelihood is between 1 to 3 \cite[this choice differs substantially from the optimal for general pseudo-marginal MCMC algorithms, see][]{sherlock2016pseudo}.

Recent theoretical work has shown that introducing correlation into the estimates of the likelihood across successive iterations can substantially improve mixing \cite[]{Deligiannidis:2015,murray2016pseudo}. For particle MCMC the idea would
then be to couple the randomness in the resampling and propagation steps of the particle filter, so that two successive runs of the filter, with similar parameter values, would generate similar particles
and trajectories and hence similar estimates of the likelihood. This would reduce the variance in the ratio of likelihood estimates that appear in the acceptance probability, and hence improve the acceptance rate.
Simulating such coupled particle filters is challenging, but see \cite{Sen/Thierry/Jasra:2016} and \cite{Jacob/Lindsten/Schon:2016} for recent approaches.

\subsubsection{Particle Gibbs}

An alternative particle MCMC algorithm is based around using a particle filter to approximate a Gibbs update. Our target distribution is the joint posterior for the parameter, $\btheta$, and the trajectory of 
the state, $\bx_{0:n}$. A Gibbs sampler would iterate between updating $\btheta$ from its full conditional given $\bx_{0:n}$, and then simulating $\bx_{0:n}$ given $\btheta$. For many models the former update is relatively
simple to perform, whereas the latter is intractable. 
A Gibbs algorithm that updates only one component $\bx_t$ at a time by
Metropolis-Hastings steps would be feasible, but convergence is usually
slow. For some models, data augmentation has been used successfully 
\cite[see, e.g.][]{Fruehwirth:09}, but it is often restricted to models
with specific structure. The idea of particles Gibbs is to use a particle
filter as a generic way of updating the whole path, $\bx_{0:n}$. Whilst the
particle filter only samples from an approximation to the true conditional
distribution of the path given $\theta$ and $\by_{1:n}$, the particle Gibbs
algorithm is designed such that the resulting MCMC algorithm still has the
true posterior as its stationary distribution.

The particle Gibbs sampler updates $\bx_{0:n}$ by drawing from a
distribution that depends not only on the current value $\btheta$, but also
on the current value of the state sequence, denoted $\bx^{cur}_{0:n}$.
Viewed as such, the term particle Gibbs is, in fact, a slight misnomer -- though it can be viewed as a Gibbs sampler on an extended state space. Drawing 
$\bx_{0:n}$ given $\btheta$ and $\bx^{cur}_{0:n}$ is based on running
a {\em conditional particle filter} algorithm such that 
one of the final particles has a trajectory that is identical to 
$\bx_{0:n}^{cur}$. In particular this means that we only ever simulate $N-1$ particles at each iteration of the conditional particle filter,
as the remaining particle is set to the corresponding entry of 
$\bx_{0:n}^{cur}$. 

The conditional particle filter algorithm is as follows
\begin{itemize}
 \item[(0)] Condition on the current state trajectory $\bx_{0:n}^{cur}$,
   and parameter value, $\btheta$. Set $\bx_0^1=\bx_0^{cur}$ and
   independently simulate $\bx_0^i$ from $\pi_0(d\bx_0\mid \btheta)$ for
   $i=2,\ldots,N$. Set $t=1$. 
 \item[(1)] Resample: If $t>1$ set $\tilde{\bx}_{t-1}^i=\bx_{t-1}^{A(i)}$ where $A(1)=1$ and $\pr(A(i)=j)=w_{t-1}^j$ for $i=2,\ldots,N$.
 \item[(2)] Propagate: Set $\bx_t^1=\bx_t^{cur}$ and draw $\bx_t^i$ from $P_{\btheta}(d\bx_t\mid \tilde{\bx}_{t-1}^i)$ for $i=2,\ldots,N$.
 \item[(3)] Reweight: Set $w_t^i \propto g_{\btheta}(\by_t\mid \bx_t^i)$. If $t<n$, set $t=t+1$ and go to step (1).
 \item[(4)] Sample and output a particle, and its associated trajectory, at
   time $n$; with the probability of sampling particle $i$ being $w_n^i$
   for $i=1\ldots,N$.  
\end{itemize}
Steps (0) to (3) are  similar to the bootstrap filter of Section
\ref{S:BootstrapFilter}, except that at each iteration we fix the first
particle to be the  corresponding part of the trajectory we are conditioning 
on. Thus at time $n$ the trajectory of the first
particle will be $\bx_{0:n}^{cur}$, and there is a non-zero probability
that the current trajectory does not change in the update. The
dynamics 
of the filter depends only on the parameter value, and this is made explicit within the notation for steps (2) and (3). 
See \cite{Andrieu/Doucet/Holenstein:2010} for a proof that updating the state trajectory using a conditional particle filter 
update leaves $p(\btheta,\bx_{0:n}\mid \by_{1:n})$ invariant. Example output from the conditional particle filter is given 
in Figure \ref{Fig:CSMC}.

\begin{figure}
 \begin{center}
  \includegraphics[scale=0.8]{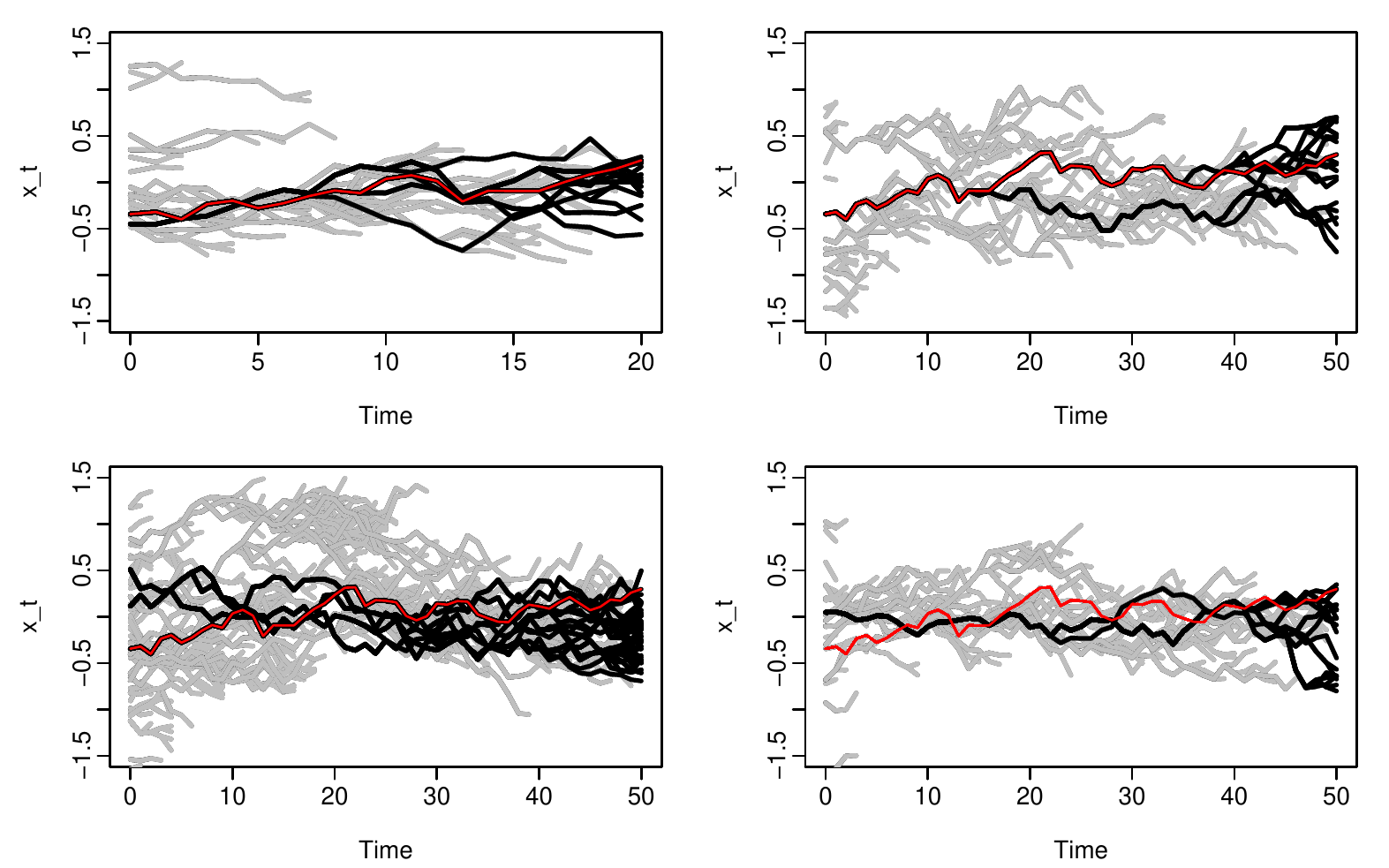}
 \end{center}
\caption{\label{Fig:CSMC} 
Output from the conditional particle filter  for the stochastic volatility model. For each figure the conditioned path is in red, the 
paths of particles from intermediate steps of the filter are in grey, and the paths associated with the particles at the final time-step are in black. Top-left plot is for $n=20$ observations and $N=30$ particles; 
bottom-left is for $n=50$ and $N=75$, right-hand plots are for $n=50$ and $N=30$, but for the bottom-right plot we have used ancestor sampling. 
The main issue with the conditional particle filter is that it can struggle to change the start of the state trajectory unless sufficient particles are used: compare top-right and bottom-left plots. A rule of thumb
is that we need to increase $N$ linearly with any increase in $n$ -- hence the similar behaviour we obtain for the two left-hand plots.  
The use of ancestor sampling can overcome this problem -- compare the two right-hand plots. In particular the sample path of the particles in the bottom-right plot is different from 
that of the path we conditioned on due to the ancestor sampling.
}
\end{figure}

As with
particle Metropolis Hastings algorithms, empirical and theoretical results
suggest that as the number of observations, $n$, increases we would need to increase the number of particles, $N$, linearly to maintain a fixed level
of mixing of the resulting MCMC algorithm. The conditional particle filter update is more efficient at updating later values of the state than earlier ones, a property that is linked to the sample impoverishment
of the simple smoothing algorithm where we store the trajectory of the particles in the filter (see the discussion in Section \ref{S:ParameterEstimation}), and that can be seen from the top row of Figure \ref{Fig:CSMC}. 
However, there are very strong results on the mixing of particle Gibbs if
sufficiently many particles are used \cite[]{Chopin/Singh:2015,Lindsten:2015,DelMoral:2016,andrieu2013uniform}.

There have been a number of extensions to the particle Gibbs sampler. First, most improvements on the bootstrap filter can be applied to the conditional particle filter sampler.
For example, balanced resampling can be used instead of multinomial resampling, and this improves mixing \cite[]{Chopin/Singh:2015}. However care is needed, as resampling in step (1) above needs to be 
from the conditional distribution of the balanced resampling scheme given that particle 1 must have at least 1 offspring \cite[see][Appendix A for more details]{Andrieu/Doucet/Holenstein:2010}.
It is also possible to extend the conditional particle filter update so as to use better proposal distributions, as in the auxiliary particle filter. 

Second, it is possible to employ ideas from the forward-backward smoother to increase the mixing of the trajectory at early time-points \cite[]{whiteley2010discussion}. The idea here is that, in step (4), rather than simulating a trajectory associated with one
of the $N$ particles at time $n$ we can use backward simulation to obtain a new trajectory given all the particles that have been stored at time $0$ to $n$. As we simulate a single trajectory, the cost of the backward 
simulation will just be linear in $N$. A particular implementation of this idea has been termed ancestor sampling \cite[]{Lindsten:2014}. After each iteration of the conditional particle filter algorithm
they sample a new ancestor for the first particle, i.e. the particle from the conditioned path. This means that whilst there is still degeneracy of the paths in the conditional particle filter, this degenerate path
is different from the conditioned path, and hence we get better mixing in the actual MCMC algorithm -- see Figure \ref{Fig:CSMC} for an example. For theoretical support for such algorithms see \cite{Chopin/Singh:2015}.
It is also possible to use a conditional particle filter to update blocks of the trajectory, rather than the whole trajectory, and this can lead to an algorithm whose computational
cost scales linearly, rather than quadratically, with the number of observations \cite[]{Singh/Lindsten/Moulines:2015}.

Third, even if the conditional particle filter update mixes well, the resulting particle Gibbs algorithm can be poor if there is strong dependence between the parameter and the trajectory. 
It is possible to overcome this by performing partial updates of the parameters within the conditional particle filter update \cite[see][]{Fearnhead/Meligkotsidou:2016}.

\section{SUMMARY AND OUTLOOK}


The following points contain the main messages of this review.

\begin{summary}[SUMMARY POINTS]
 \begin{enumerate}
 \item Filtering and data assimilation combine partial observation with
a dynamical model to estimate latent states of a system.
 \item Particle filters are completely general, but often suffer from
sample depletion. Ensemble Kalman filters are more robust,
but rely on Gaussian assumptions.
 \item Combinations of particle and MCMC methods are promising new
developments for joint estimation of parameters and states.
 \end{enumerate}
 \end{summary}

 \begin{issues}[FUTURE ISSUES]
 \begin{enumerate}
 \item  How do we best exploit the differents strengths of particle and ensemble Kalman
filters to improve filtering of high-dimensional system with
nonlinear and non-Gaussian features? Are there version of particle filters than can move particles,
like the ensemble Kalman filter does, instead of re-weighting them?
 \item There is a need to develop theory for better understanding of localized ensemble Kalman filters
in realistic settings where the number of particles is much smaller
than the dimension of the state.
 \item How do we design particle filter and related algorithms to best take advantage of modern computer architectures?
 \end{enumerate}
 \end{issues}

\section*{DISCLOSURE STATEMENT}
The authors are not aware of any affiliations, memberships, funding, or
financial holdings that 
might be perceived as affecting the objectivity of this review. 

\section*{ACKNOWLEDGMENTS}
We are grateful to Sylvain Robert for producing Fig.\ref{Fig:enkf}  and
the animations of the Lorenz 96 model in the supplemental material. We would like to thank
Christophe Andrieu, Nicolas Chopin, Sylvain Robert, Chris Sherlock and an anonymous reviewer for helpful comments
on an earlier version of this review article.
Paul Fearnhead was funded by the EPSRC programme grant EP/K014463 (\em{i-like}).
%


\bibliographystyle{ar-style1}
\bibliography{PFreview}

\end{document}